\documentclass[
a4paper,reqno]{amsart}
\usepackage[T1]{fontenc}
\usepackage[english]{babel}
\usepackage{amssymb,bbm,bm,enumerate}
\usepackage{color}
\usepackage[linktocpage=true,colorlinks=true, linkcolor=blue, citecolor=red, urlcolor=green]{hyperref}
\usepackage[all]{xy}

\usepackage{float}
\restylefloat{table}

\usepackage{tikz-cd}

\renewcommand{\ker}{\Ker}

\newcommand{\mc}[1]{\mathcal{#1}}
\newcommand{\mf}[1]{\mathfrak{#1}}
\newcommand{\mb}[1]{\mathbb{#1}}

\newcommand{\ul}[1]{\underline {#1}}
\newcommand{\id}{\mathbbm{1}}
\newcommand{\quot}[2] {\ensuremath{\raisebox{.40ex}{\ensuremath{#1}}
\! \big / \! \raisebox{-.40ex}{\ensuremath{#2}}}}
\newcommand{\tint}{{\textstyle\int}}
\newcommand{\assr}[1]{\stackrel{#1}{\longrightarrow}}
\newcommand{\assl}[1]{\stackrel{#1}{\longleftarrow}}
\newcommand{\as}{\mathop{\rm as }}
\newcommand{\ldb}{\{\!\!\{}
\newcommand{\rdb}{\}\!\!\}}

\DeclareMathOperator{\Mat}{Mat}

\DeclareMathOperator{\ad}{ad}
\DeclareMathOperator{\im}{Im}

\DeclareMathOperator{\Ker}{Ker}
\DeclareMathOperator{\Span}{Span}

\DeclareMathOperator{\Res}{Res}

\DeclareMathOperator{\mult}{m}
\DeclareMathOperator{\Tr}{Tr}

\theoremstyle{plain}

\theoremstyle{definition}

\theoremstyle{remark}

\numberwithin{equation}{section}

\definecolor{light}{gray}{.9}

\begin{document}

\title{Poisson vertex algebras and Hamiltonian PDE}

\author{Alberto De Sole}
\address{Dipartimento di Matematica, Sapienza Universit\`a di Roma,
P.le Aldo Moro 5, 00185 Rome, Italy}
\email{desole@mat.uniroma1.it}
\urladdr{www1.mat.uniroma1.it/\$$\sim$\$desole}
\author{Victor G. Kac}
\address{Dept of Mathematics, MIT,
77 Massachusetts Avenue, Cambridge, MA 02139, USA}
\email{kac@math.mit.edu}
\author{Daniele Valeri}
\address{Dipartimento di Matematica, Sapienza Universit\`a di Roma,
P.le Aldo Moro 5, 00185 Rome, Italy}
\email{daniele.valeri@uniroma1.it}



\begin{abstract}
%
We discuss the theory of Poisson vertex algebras and their generalizations
in relation to integrability of Hamiltonian PDE.
In particular, we discuss the theory of affine classical $\mc W$-algebras
and apply it to construct a large class of integrable Hamiltonian PDE.
We also discuss non commutative Hamiltonian PDE
in the framework of double PVA,
and differential-difference Hamiltonian equations
in the framework of multiplicative PVA.
\end{abstract}

%
%

\maketitle

\tableofcontents

\section{Introduction}
\label{sec:1}

Poisson algebras naturally appear as the classical limit of a sufficiently nice family of associative algebras
$A_\hbar$, $\hbar$ in the base field $\mb F$. Namely, if the algebra $A_0$ is commutative and $[A_{\hbar},A_{\hbar}]\subset \hbar A_{\hbar}$,
then the factor algebra $\mc A= A_\hbar/\hbar A_\hbar$ is a commutative
associative algebra, endowed with a Lie algebra bracket
\begin{equation}\label{intro1}
\{a, b\}=\lim_{\hbar\to0}\frac1\hbar[\tilde a,\tilde b]\,,
\end{equation}
where $\tilde a$ and $\tilde b$ are some preimages of $a$ and $b$ under the canonical map
$A_\hbar\to\mc A$
and $[\cdot\,,\,\cdot]$ denotes the commutator in $A_\hbar$.
It is easy to see that \eqref{intro1} is well defined and that the product and the bracket of $\mc A$ satisfy the Leibniz rules:
\begin{equation}\label{intro2}
\{a,bc\}=\{a,b\}c+b\{a,c\}\,,
\qquad
\{ab,c\}=\{a,c\}b+a\{b,c\}\,.
\end{equation}
Thus $\mc A$ is a Poisson algebra, which is defined as a 
commutative associative algebra,
endowed with a Lie algebra bracket satisfying the Leibniz rules \eqref{intro2}.

Poisson algebras play an important role in the theory of Hamiltonian ODE. Namely, if
$\mc P$ is the algebra of smooth functions on a smooth manifold $X$, endowed with a Poisson bracket
$\{\cdot\,,\,\cdot\}$, then for each $h\in \mc P$ the corresponding Hamiltonian equation is defined as
\begin{equation}\label{intro3}
\frac{df}{dt}=\{h,f\}\,,\qquad f\in\mc P\,.
\end{equation}
By Liouville theory, equation \eqref{intro3} can be solved ``by quadratures'' if $h$ can be included in a ``sufficiently large''
commutative (with respect to the Poisson bracket) subalgebra of $\mc P$.
The elements of this subalgebra are called integrals of motion of the equation \eqref{intro3}.

In local coordinates $x=(x_i)_{i=1}^d$ of $X$ the Poisson bracket is defined, using the matrix
$H(x)=\left(\{x_j,x_i\}\right)_{i,j=1}^d$, by the ``Master Formula''
\begin{equation}\label{intro4}
\{f,g\}=\frac{\partial g}{\partial x}\cdot H(x)\frac{\partial f}{\partial x}\,,
\end{equation}
where $\frac{\partial f}{\partial x}$ is the column vector of the partial derivatives $\frac{\partial f}{\partial x_i}$.

In a similar way, Poisson vertex algebras (PVA) appear in the theory of vertex algebras. The difference is that
the classical limit in this case is a unital commutative associative differential algebra with a derivation $\partial$,
endowed with a Lie conformal algebra $\lambda$-bracket $\{a_\lambda b\}$ instead of a Lie algebra bracket $\{a,b\}$.

As explained in Section \ref{sec:2}, the $\lambda$-bracket should satisfy the Lie conformal algebra 
axioms \eqref{sesqui}, \eqref{skewsim}
and \eqref{jacobi}, and the Leibniz rules \eqref{lleibniz}, \eqref{rleibniz}.

As we shall explain, the PVA $\mc V$ play an important role in the theory of Hamiltonian  PDE. The key differences 
with the Hamiltonian ODE are as follows.
First, equation \eqref{intro3} is replaced by
$$
\frac{df}{dt}=\{h_\lambda f\}|_{\lambda=0}\,,
\quad
f\in\mc V\,.
$$
Second, this evolution depends on the Hamiltonian functional $\tint h\in \mc V/\partial\mc V$,
where $\tint:\mc V\to\mc V/\partial\mc V$ is the canonical map,
and $\mc V/\partial\mc V$ is endowed with the Lie algebra bracket
$$
\{\tint f,\tint g\}=\tint \{f_\lambda g\}|_{\lambda=0}\,.
$$
Moreover, the integrals of motion are elements of an abelian subalgebra of the Lie algebra $\mc V/\partial\mc V$
which contains $\tint h$.
Finally, the ``Master Formula'' is more complicated than \eqref{intro4} (see \eqref{masterformula}),
and leads to the following formulas in local coordinates $u_i(t)$, which are analogues of \eqref{intro3} for $f=x_i$
and \eqref{intro4} respectively
\begin{equation}\label{intro7}
\frac{du}{dt}=H(\partial)\frac{\delta\tint h}{\delta u}\,,
\end{equation}
\begin{equation}\label{intro8}
\{\tint f,\tint g\}=\int \frac{\delta\tint g}{\delta u} \cdot H(\partial)\frac{\delta\tint f}{\delta u}
\,.
\end{equation}
Here $u$ is the column vector of the $u_i$'s, $H(\partial)$ is the matrix differential operator with entries
$\{{u_j}_\partial u_i\}_\to$, and $\frac{\delta\tint f}{\delta u}$ is the column vector of variational derivatives 
$\frac{\delta f}{\delta u_i}=\sum_{n\geq0}(-\partial)^n\frac{\partial f}{\partial u_i^{(n)}}$
(elements from $\partial\mc V$ have zero variational derivatives).

Note that, while the theory of Hamiltonian ODE is several centuries old, the basics \eqref{intro7} and \eqref{intro8}
of the theory of Hamiltonian PDE originated in the works of Gardner and Faddeev-Zakharov from early 1970's.

In the present article we review the developments of the theory of Hamiltonian PDE in this century, based on the
``coordinate free'' approach of Poisson vertex algebras.

Our list of references is far from being complete.
We refer the reader to references in the quoted papers for further information.

\smallskip

{\bf Authors note:} this article is the preprint version of our contribution to the book titled "Encyclopedia of Mathematical Physics 2nd edition".

\section{PVA and Master Formula}
\label{sec:2}

By a differential algebra we mean a unital commutative associative algebra
$\mc V$ over a field $\mb F$ of characteristic $0$, with a derivation $\partial:\,\mc V\to\mc V$.
A \emph{$\lambda$-bracket} on a differential algebra $\mc V$ is an $\mb F$-linear map
$\mc V\otimes\mc V\to\mb F[\lambda]\otimes\mc V$,
denoted by $f\otimes g\to\{f_\lambda g\}$,
satisfying \emph{sesquilinearity} ($f,g\in\mc V$):
\begin{equation}\label{sesqui}
\{\partial f_\lambda g\}=
-\lambda\{f_\lambda g\},\qquad 
\{f_\lambda\partial g\}=
(\lambda+\partial)\{f_\lambda g\}\,,
\end{equation} 
and the \emph{left and right Leibniz rules} ($f,g,h\in\mc V$):
\begin{align}\label{lleibniz}
\{f_\lambda gh\}&=
\{f_\lambda g\}h+\{f_\lambda h\}g,\\
\{fh_\lambda g\}&=
\{f_{\lambda+\partial}g\}_{\rightarrow}h+\{h_{\lambda+\partial}g\}_{\rightarrow}f\,,\label{rleibniz}
\end{align}
where we use the following notation: if
$\{f_\lambda g\}=\sum_{n\in\mb Z_{\geq0}}\lambda^n c_n$,
then
$\{f_{\lambda+\partial}g\}_{\rightarrow}h=\sum_{n\in\mb Z_{\geq0}}c_n(\lambda+\partial)^nh$.
We say that the $\lambda$-bracket is \emph{skew-symmetric} if
\begin{equation}\label{skewsim}
\{g_\lambda f\}=-\{f_{-\lambda-\partial}g\}\,,
\end{equation}
where, now, 
$\{f_{-\lambda-\partial}g\}=\sum_{n\in\mb Z_{\geq0}}(-\lambda-\partial)^nc_n$
(if there is no arrow we move $\partial$ to the left).
Note that the right Leibniz rule \eqref{rleibniz} follows from the left Leibniz rule \eqref{lleibniz}
and skew-symmetry \eqref{skewsim}.

A \emph{Poisson vertex algebra} (PVA) is a differential algebra $\mc V$ endowed 
with a $\lambda$-bracket which is skew-symmetric and satisfies 
the following \emph{Jacobi identity} in $\mc V[\lambda,\mu]$ ($f,g,h\in\mc V$):
\begin{equation}\label{jacobi}
\{f_\lambda\{g_\mu h\}\}=
\{\{f_\lambda g\}_{\lambda+\mu}h\}+
\{g_\mu\{f_\lambda h\}\}\,.
\end{equation}

In the theory of integrable systems the most important example of a differential algebra is
the \emph{algebra of differential polynomials} in $\ell$ variables:
$$
\mc V_\ell=\mb F[u_i^{(n)}\mid i\in I=\{1,\ldots,\ell\},n\in\mb Z_{\geq0}]\,,
$$
where $\partial$ is the derivation defined by $\partial(u_i^{(n)})=u_i^{(n+1)}$, $i\in I,n\in\mb Z_{\geq0}$.
Note that in $\mc V_\ell$ we have the following commutation relations:
\begin{equation}\label{partialcomm}
\left[\frac{\partial}{\partial u_i^{(n)}},\partial\right]=\frac{\partial}{\partial u_i^{(n-1)}}\,,
\end{equation}
where the RHS is considered to be zero if $n=0$.
More general examples are \emph{algebras of differential functions}, which are differential algebra 
extensions of the algebra $\mc V_\ell$, with commuting partial 
derivatives $\frac{\partial}{\partial u_i^{(n)}}$ satisfying \eqref{partialcomm} and such that
for every $f\in\mc V$ we have $\frac{\partial f}{\partial u_i^{(n)}}=0$ for all but finitely many values 
of $i\in I$ and $n\in\mb Z_{\geq0}$.

If $\mc V$ is an algebra of differential functions in the variables $\{u_i\}_{i\in I}$ 
and $\big(H_{ji}(\lambda)\big)_{i,j\in I}$ is an $\ell\times\ell$-matrix with coefficients in $\mc V[\lambda]$,
one gets a well defined $\lambda$-bracket on $\mc V$ by the following Master Formula \cite[Sec. 6]{DSK}:
\begin{equation}\label{masterformula}
\{f_\lambda g\}=
\sum_{\substack{i,j\in I\\m,n\in\mb Z_{\geq0}}}\frac{\partial g}{\partial u_j^{(n)}}(\lambda+\partial)^n
H_{ji}(\lambda+\partial)(-\lambda-\partial)^m\frac{\partial f}{\partial u_i^{(m)}}
\end{equation}
In particular, 
\begin{equation}\label{masterformula2}
\{u_i{}_\lambda u_j\}=H_{ji}(\lambda),\,i,j\in I
\,.
\end{equation}
Moreover, if $\mc V=\mc V_\ell$ is the algebra of differential polynomials,
\eqref{masterformula}-\eqref{masterformula2} necessarily hold.
It is explained in \cite[Sec. 1]{BDSK09} that, like in the Poisson algebra case,
the necessary and sufficient conditions for validity of PVA
axioms on $\mc V$ are the skew-symmetry for each pair $u_i,u_j$ 
and the Jacobi identity for each triple $u_i,u_j,u_k$.
Skew-symmetry is equivalent to the skew-adjointness of the matrix differential
operator $H(\partial)=\big(H_{ji}(\partial)\big)_{i,j\in I}$,
while Jacobi identity is equivalent to $[H(\partial),H(\partial)]=0$,
where $[\cdot\,,\,\cdot]$ denotes the Schouten-bracket.

It is shown in \cite{Ar12} that any Poisson algebra $\mc P$ can be naturally extended to yield a structure of a PVA on its jet algebra $J_{\infty}(\mc P)$ (which is the universal differential algebra associated with $\mc P$) such that
$$
\{f_\lambda g\}=\{f,g\}\,,
$$
for every $f,g\in\mc P\subset J_{\infty}(\mc P)$.

Another important example is  the \emph{affine PVA} $\mc V(\mf g)$ associated to a finite dimensional Lie algebra $\mf g$
with a non-degenerate symmetric invariant bilinear form $(\cdot\,|\,\cdot)$. As a differential algebra it
is $\mc V(\mf g)=S(\mb F[\partial]\mf g)$, the algebra of differential polynomials in a basis of $\mf g$. A family of PVA
structures is given, on generators $a,b\in\mf g$, by
\begin{equation}\label{lambda_affine}
\{a_\lambda b\}_z=[a,b]+(a|b)\lambda + z (s|[a,b])
\,,\qquad z\in\mb F\,,
\end{equation}
where $s\in\mf g$ is a fixed element, and it is extended using sesquilinearity and Leibniz rules.

\section{Exactness of the variational complex}
\label{sec:3}

For $f\in\mc V$, we denote by $\tint f$
the image of $f$ in the quotient space $\mc V/\partial\mc V$.
The sign of integral is used because, in the quotient space $\mc V/\partial\mc V$
integration by parts holds:
$\tint(\partial f)g=-\tint f(\partial g)$.
Let $\mc V$ be an algebra of differential functions in the variables $u_i$, $i\in I$.
By \eqref{partialcomm} we have a well-defined \emph{variational derivative}
$\frac{\delta}{\delta u}:\,\mc V/\partial\mc V\to\mc V^{\ell}$,
which is the column of variational derivatives
$$
\frac{\delta\tint f}{\delta u_i}=\sum_{n\in\mb Z_{\geq0}}(-\partial)^n\frac{\partial f}{\partial u_i^{(n)}}\,\,,\,\,\,\, i\in I\,.
$$
(We will often simply write $\frac{\delta f}{\delta u_i}$ in place of $\frac{\delta\tint f}{\delta u_i}$ as the variational derivative does not depend on the choice of the representative).

For $X\in\mc V^m$, $m\geq1$,
we define its \emph{Frechet derivative} as the differential operator
$D_X(\partial):\,\mc V^\ell\to\mc V^m$ given by
\begin{equation}\label{20111020:eq1}
\big(D_X(\partial)P\big)_j
=\sum_{n\in\mb Z_{\geq0}}\sum_{i\in I}\frac{\partial X_j}{\partial u_i^{(n)}} \partial^n P_i
\,\,,\,\,\,\,j=1,\dots,m
\,.
\end{equation}

The above notions are linked naturally in the variational complex:
$$
0\to\mc V/\partial\mc V
\stackrel{\frac{\delta}{\delta u}}{\longrightarrow}
\mc V^{\ell}
\stackrel{\delta}{\longrightarrow}
\Sigma_\ell
\to\dots
$$
where $\Sigma_\ell$ is the space of skewadjoint $\ell\times\ell$ 
matrix differential operators over $\mc V$,
and $\delta(F)=D_F(\partial)-D^*_F(\partial)$,
for $F\in\mc V^{\ell}$.
The construction of the whole complex can be found in \cite{BDSK09,DSK09}.
It is proved there that the variational complex
is exact, provided that the algebra of differential functions $\mc V$ is normal,
i.e. for each $i\in I$, $n\in\mb Z_{\geq0}$,
$\frac{\partial}{\partial u_i^{(n)}}$ is surjective on 
$\displaystyle{\mc V_{i,n}:=\bigcap_{(j,k)>(i,n)}\ker\Big(\frac{\partial}{\partial u_j^{(k)}}\Big)}$.
(For example, the algebra $\mc V_\ell$ of differential polynomials is normal, and, clearly, 
any algebra of differential functions can be embedded in a normal one.)
In particular, if $\mc V$ is normal, we have 
$$
\ker\big(\frac\delta{\delta u}\big)=\mc F+\partial\mc V
\,,
$$
where 
$$
\mc F=\Big\{f \in\mc V \Big| \frac{\partial f}{\partial u_{i}^{(n)}}=0\,, \text{ for all }i\in I,n\in\mb Z_{\geq0}\Big\}
\,,
$$
and that $F\in\mc V^{\ell}$ is \emph{closed}, i.e. its Frechet derivative $D_F(\partial)$ is selfadjoint,
if and only if it is \emph{exact}, i.e. $F\in\frac{\delta}{\delta u}(\mc V/\partial\mc V)$.

\section{Translation of the GFZ formalism to \texorpdfstring{$\lambda$}{lambda}-bracket formalism}
\label{sec:4}
For a PVA $\mc V$,
we call $\quot{\mc V}{\partial\mc V}$ the space of \emph{local functionals}.
It follows from the sesquilinearity axioms \eqref{sesqui} that
\begin{equation}\label{20230110:eq1}
\{\tint g,\tint h\}:=\tint\{g_\lambda h\}\big|_{\lambda=0}\in\quot{\mc V}{\partial\mc V}
\,\,\text{ and }\,\,
\{\tint g,h\}:=\{g_\lambda h\}\big|_{\lambda=0}\in\mc V
\,,
\end{equation}
are well defined.
It follows from skewsymmetry \eqref{skewsim} and the Jacobi identity axioms \eqref{jacobi} of a PVA
that $\{\tint f,\tint g\}$ is a Lie algebra bracket on $\quot{\mc V}{\partial\mc V}$,
and that $\{\tint f,g\}$ is a representation of  $\quot{\mc V}{\partial\mc V}$ on $\mc V$
by derivations of the associative product on $\mc V$ commuting with $\partial$.

The \emph{Hamiltonian equation} associated 
to a \emph{Hamiltonian functional} $\tint h\in\quot{\mc V}{\partial\mc V}$
is, by definition,
\begin{equation}\label{20130314:eq1}
\frac{du}{dt}=\{\tint h,u\}
\,\,,\,\,\,\,
u\in\mc V\,.
\end{equation}
An \emph{integral of motion} of equation \eqref{20130314:eq1} is 
a local functional $\tint g\in\quot{\mc V}{\partial\mc V}$
such that $\{\tint h,\tint g\}=0$,
in which case one says that $\tint h,\tint g$ are \emph{in involution}.
Equation \eqref{20130314:eq1} is called \emph{integrable}
if there are infinitely many linearly independent integrals of motion in involution,
$\tint h_n,\,n\in\mb Z_{\geq0}$, where $h_0=h$.
In this case, we have an integrable hierarchy of Hamiltonian equations
\begin{equation}\label{20130314:eq1b}
\frac{du}{dt_n}=\{\tint h_n,u\}
\,\,,\,\,\,\,
u\in\mc V\,,n\in\mb Z_{\geq0}\,.
\end{equation}

The notion of a PVA appears naturally in the theory of vertex algebras as their classical limit, cf. \cite[Sec. 6]{DSK},
in the same way as a Poisson algebra appears naturally as a classical limit of a family of associative algebras.

However, the theory of Hamiltonian PDEs was started 15 years before the advent of the vertex algebra theory,
in the work of Faddeev and Zakharov \cite{ZF71}, who attribute the construction to C. S. Gardner, which, in algebraic terms,
is as follows.
Let, as in Section \ref{sec:2}, $\mc V=\mc V_\ell$ be the algebra of differential polynomials in $\ell$ differential
variables.
Given $\tint h\in\mc V/\partial\mc V$,
one defines a Hamiltonian PDE
\begin{equation}\label{intro:eq3}
\frac{du}{dt}=H(\partial)\frac{\delta }{\delta u}\tint h\,,
\end{equation}
where $u=(u_i)_{i\in I}$ is the $\ell$-column vector of differential variables,
$H(\partial)$ is an $\ell\times\ell$-matrix differential operator with coefficients
in $\mc V$, called the \emph{Poisson structure} of $\mc V$,
and $\frac{\delta}{\delta u}\tint h=\big(\frac{\delta}{\delta u_i}\tint h\big)_{i\in I}$ is the $\ell$-column vector
of variational derivatives.

The basic assumption on the Poisson structure $H(\partial)$ is that formula
\begin{equation}\label{intro:eq4}
\{\tint f,\tint g\}=\int \frac{\delta \int g}{\delta u}H(\partial)\frac{\delta \int f}{\delta u}
\end{equation}
defines a Lie algebra structure on $\mc V/\partial\mc V$.
A simple observation is that equation \eqref{intro:eq3} coincides with \eqref{20130314:eq1},
and \eqref{intro:eq4} coincides with the first equation in \eqref{20230110:eq1},
if we let $H(\partial)=\left(\{{u_j}_\partial u_i\}_\rightarrow\right)_{i,j\in I}$,
where the arrow means that $\partial$ should be moved to the right.
This observation follows from the Master Formula \eqref{masterformula}
and integration by parts.
It is proved in \cite[Sec. 1]{BDSK09} that \eqref{intro:eq4} satisfies the Lie algebra axioms
if and only if \eqref{masterformula} defines a structure of a PVA on $\mc V$.
Thus PVAs provide a coordinate free approach to the theory of Hamiltonian PDEs.

The simplest non-trivial example of a Hamiltonian PDE is the celebrated KdV equation
\begin{equation}\label{intro:eq5}
\frac{du}{dt}=3uu'+cu'''\,,\quad c\in\mb F
\,,
\end{equation}
the only one studied in \cite{ZF71}.
It is Hamiltonian for the algebra of differential polynomials in one variable $\mc V=\mb F[u,u',u'',\dots]$,
the local Hamiltonian functional $\tint h=\frac12\tint u^2$, and the $\lambda$-bracket, defined
on the generator $u$ by
$$
\{u_\lambda u\}=(\partial+2\lambda)u+c\lambda^3\,,
$$
and uniquely extended to $\mc V$ by the PVA axioms,
which corresponds to the Poisson structure
\begin{equation}\label{intro:eq7}
H(\partial)=u'+2u\partial+c\partial^3\,,
\end{equation}
called the Virasoro-Magri Poisson structure.
In fact, equation \eqref{intro:eq5} is Hamiltonian also for another choice of the local Hamiltonian functional and the
Poisson structure: $\tint h_2=\frac12\tint(u^3+cuu'')$, $H_0(\partial)=\partial$ (known as the Gardner-Fadeev-Zakharov (GFZ) Poisson structure), which makes the KdV
equation a bi-Hamiltonian PDE (see \cite{Mag78}).

Gardner, Green, Kruskal and Miura proved in \cite{GGKM67} that the KdV equation \eqref{intro:eq5}
admits infinitely many linearly independent integrals of motion, which initiated the whole theory of integrable PDEs.
It follows from \cite{ZF71} that these integrals are in involution with respect to the bracket \eqref{intro:eq4}, where
$H(\partial)$ is given by \eqref{intro:eq7}. This leads to the whole KdV hierarchy as in \eqref{20130314:eq1b}.

\section{Lenard-Magri scheme}
\label{sec:5}

In the paper \cite{Mag78} Magri proposed a simple algorithm, called nowdays the \emph{Lenard-Magri scheme},
which allows one to prove that integrals of motion of a Hamiltonian PDE \eqref{intro:eq3} are in involution,
provided that the same equation can be written in two compatible Hamiltonian forms,
as it happens, for example, for the KdV equation.

We describe this algorithm in the PVA framework. Let $\{\cdot\,_\lambda\,\cdot\}_0$ and $\{\cdot\,_\lambda\,\cdot\}_\infty$
be two $\lambda$-brackets on the same differential algebra $\mc V$.
We can consider the pencil of $\lambda$-brackets
$$
\{\cdot\,_\lambda\,\cdot\}_z
=
\{\cdot\,_\lambda\,\cdot\}_0
+
z\{\cdot\,_\lambda\,\cdot\}_\infty
\quad,\qquad z\in\mb F
\,.
$$
We say that $\mc V$ is a \emph{bi-PVA}
if $\{\cdot\,_\lambda\,\cdot\}_z$ is a PVA $\lambda$-bracket on $\mc V$ for every $z\in\mb F$.
In this case the $\lambda$-brackets $\{\cdot\,_\lambda\,\cdot\}_0$ and $\{\cdot\,_\lambda\,\cdot\}_\infty$ are called
\emph{compatible.}

Given a bi-PVA $\mc V$,
a \emph{bi-Hamiltonian} equation is an evolution equation
which can be written in Hamiltonian form with respect to both PVA $\lambda$-brackets
and two Hamiltonian functionals $\tint h_0,\tint h_1\in\mc V/\partial\mc V$:
$$
\frac{du}{dt}
=
\{\tint h_0,u\}_0
=
\{\tint h_1,u\}_\infty
\,,\,\, u\in\mc V
\,.
$$

The \emph{Lenard-Magri recurrence relation} (see \cite{Mag78}) is the following recursive equation:
\begin{equation}\label{eq:LM}
\{\tint h_n,\tint u\}_0
=
\{\tint h_{n+1},\tint u\}_\infty
\,\,,\,\,\,\,
n\in\mb Z_{\geq0}\,,\, u\in\mc V
\,.
\end{equation}
If solved, the Lenard-Magri recurrence relation \eqref{eq:LM} produces local functionals in involution:
$$
\{\tint h_m,\tint h_n\}_0
=
\{\tint h_m,\tint h_n\}_\infty
=
0
\,\,\text{ for all }\, m,n\geq0
\,,
$$
namely
$\mc A:=\Span\{\tint h_n\}_{n=0}^\infty\subset\mc V/\partial\mc V$
is an abelian subalgebra with respect to both Lie algebra brackets $\{\cdot\,,\,\cdot\}_0$ and $\{\cdot\,,\,\cdot\}_\infty$.
In this way,
we get the corresponding hierarchy of bi-Hamiltonian equations
$$
\frac{du}{dt_n}
=
\{\tint h_n,u\}_0
=
\{\tint h_{n+1},u\}_\infty
\,,\,\,
n\in\mb Z_{\geq0},\, u\in\mc V
\,,
$$
which is integrable, provided that $\mc A$ is infinite dimensional.

Let $\mc V$ be an algebra of differential functions in the variables $u_1,\dots,u_\ell$, and let us assume that
it is a bi-PVA with $\lambda$-bracket $\{\cdot\,_\lambda\,\cdot\}_z=\{\cdot\,_\lambda\,\cdot\}_{H_1}+z\{\cdot\,_\lambda\,\cdot\}_{H_0}$ corresponding to a compatible pair of Poisson structures $H_0(\partial)$ and $H_1(\partial)$.
By the Master Formula \eqref{masterformula}, the Lenard-Magri recurrence \eqref{eq:LM} reads
\begin{equation}\label{eq:LM2}
H_1(\partial)\frac{\delta h_n}{\delta u}=H_0(\partial)\frac{\delta h_{n+1}}{\delta u}\,,
\quad n\in\mb Z_{\geq0}
\,.
\end{equation}
In order to solve the recurrence \eqref{eq:LM2}, one needs to solve two problems. First, one needs to find an infinite
sequence $\{F_n\}_{n\in\mb Z_{\geq0}}\subset\mc V^\ell$, starting with $F_0=\frac{\delta h_0}{\delta u}$, $F_1=\frac{\delta h_1}{\delta u}$, such that
$$
H_1(\partial)F_n=H_0(\partial)F_{n+1}\,,
\quad n\in\mb Z_{\geq0}
\,.
$$
Then, one has to show that $F_n=\frac{\delta h_n}{\delta u}$, for all $n\in\mb Z_{\geq0}$. 
A detailed exposition on how to solve these two problems is contained in \cite[Sec. 2]{BDSK09}. In particular, there it is proved that the sequence $\{F_n\}_{n\in\mb Z_{\geq0}}$ exists provided that
\begin{equation}\label{con1}
\Span\Big\{\frac{\delta h_0}{\delta u},\frac{\delta h_1}{\delta u}\Big\}^\perp\subset\im H_0(\partial)
\,,
\end{equation}
where the orthogonal complement is with respect to the
non-degenerate pairing 
$(\cdot\,|\,\cdot):\,\mc V^\ell\times\mc V^{\ell}\to\mc V/\partial\mc V$
given by
\begin{equation}\label{pairing}
(P|\xi)=\tint P\cdot\xi\,.
\end{equation}
Then, all the elements $F_n$ are variational derivatives, provided that $H_0(\partial)$ is non-degenerate, i.e., invertible
in the ring of pseudodifferential operators, and $\mc V$ is normal.

As an example, for the KdV hierarchy discussed in Section \ref{sec:4}, we have
\begin{align*}
&\tint h_0=\tint u\,,\,\frac{\delta h_0}{\delta u}=1\,,
\quad
\tint h_1=\frac12 \tint u^2\,,\,\frac{\delta h_1}{\delta u}=u\,,
\\
&\tint h_2=\frac12 \tint (u^3+c u u'')\,,\,\frac{\delta h_2}{\delta u}=\frac32u^2+c u''\,,
\end{align*}
hence, condition \eqref{con1} trivially holds. Moreover, $H_0(\partial)=\partial$ is non-degenerate.
This implies that KdV equation \eqref{intro:eq5},
which is \eqref{intro:eq3} for $h=h_2$ and $H(\partial)=H_0(\partial)$, is integrable.


\section{Variational Poisson cohomology and integrability}
\label{sec:6}
Let $\mc V$ be a differential algebra.
We denote by $\mf g_{-1}=\mc V/\partial \mc V$ the space of local functionals,
and by $\mf g_0$ the Lie algebra of of \emph{evolutionary vector fields}, i.e. the Lie algebra of all
derivations of $\mc V$, commuting with $\partial$. Its action on $\mc V$
descends to $\mf g_{-1}$.
We also denote by $\mf g_1$ the space of all $\lambda$-brackets on $\mc V$.

The basic construction of \cite{DSK13b} is the
$\mb Z$-graded Lie superalgebra
$$
  W^{\partial ,\as} \left(\Pi \mc V\right) = \left(\Pi
    \mf g_{-1}\right) \oplus \mf g_0
     \oplus \left(\Pi \mf g_1\right) \oplus  \dots \, ,
$$
where $\mf g_{-1}$, $\mf g_0$ and $\mf g_1$ are as above and $\Pi\mf g_i$
stands for the space $\mf g_i$ with reversed parity.  For
$\tint f, \tint g \in \Pi \mf g_{-1}$, $X,Y \in \mf g_0$ and $H \in
\Pi \mf g_1$ the commutators are as follows:
\begin{align}
\nonumber
\left[\tint f, \tint g \right] &= 0 \, ,  \\
\nonumber
   \left[X, \tint f\right]  &= \tint X(f)\, , \\
\label{commvec}
    \left[X,Y\right] &=XY-YX\, , \\
\label{eq:0.5}
    \left[\{ \, .\, {}_\lambda \,. \, \}_H, \tint f\right] (g)
       &=\{ f_\lambda g \}_H\big|_{\lambda=0}\, , \\
\label{eq:0.6}
   \{ f_\lambda g \}_{[X,H]} &= X (\{ f_\lambda g \}_H)
        - \{ X (f)_\lambda g \}_H - \{ f_\lambda , X(g)\}_H\, .
\end{align}
Here and below, for $H\in\Pi\mf g_1$,
we let $\{a_\lambda b\}_H$ be the corresponding $\lambda$-bracket of $a,b\in\mc V$.
The definition of whole Lie superalgebra
$W^{\partial,\as} (\Pi \mc V)$ can be found in \cite[Sec. 7]{DSK13b}, but for applications to Hamiltonian
PDE one needs only the condition $[H,K]=0$ for $H,K \in \Pi
\mf g_1$, which is as follows $(f,g,h \in \mc V)$:
$$
  \{\{ f_\lambda g \}_{K \, \lambda +\mu} h \}_H
    \!-\! \{ f_\lambda \{ g_\mu h \}_K \}_H \!+\!
     \{ g_\mu \{ f_\lambda h \}_K\}_H \!+\! (H \leftrightarrow K)=0\,.
$$
%
%
%
This condition holds if the $\lambda$-brackets $\{\cdot\,_\lambda\,\cdot\}_H$ and $\{\cdot\,_\lambda\,\cdot\}_K$ are compatible.
 
As mentioned in Section \ref{sec:5}, the basic device for proving integrability of a Hamiltonian
equation is the Lenard-Magri scheme \eqref{eq:LM}.
Note that, in \eqref{eq:LM}, we do not need to assume that the
$\lambda$-brackets are PVA nor that they are compatible.
These assumptions enter when we try to prove the existence of the infinite
sequence, $\tint h_n$, satisfying \eqref{eq:LM}, as we explain below.

Indeed, let $H,K \in \Pi \mf g_1$ be two compatible PVA
$\lambda$-brackets on $\mc V$.  Since $[K,K] =0$, it follows that $(\ad K)^2=0$,
hence we may consider the {\em variational cohomology complex}
$(W^{\partial,\as} (\Pi \mc V) = \bigoplus^\infty_{j=-1} W_j, \ad K )$,
where $W_{-1}=\Pi \mf g_{-1}$, $W_0 = \mf g_0$, $W_1 = \Pi \mf g_1,
\dots$.  By definition, $X \in W_0$ is \emph{closed}
if, in view of \eqref{eq:0.6}, it is a derivation of
the $\lambda$-bracket $\{ \, .\, {}_\lambda \,. \, \}_K$,
and it is exact if, in view of \eqref{eq:0.5}, $X = \{ h_\lambda \cdot \}_K
\big|_{\lambda =0}$ for some $h \in W_{-1}$.  Now we can find a
solution to \eqref{eq:LM}, by induction on $n$ as follows.  By
Jacobi identity in $W^{\partial,\as}(\Pi \mc V)$ we have $ [K,[H,h_n]] =
-[H, [K,h_n]]$, which, by inductive assumption,
equals $-[H,[H,h_{n-1}]]=0$,
since $[H,H]=0$ and $H\in W_1$ is odd.
Thus, the element $[H,h_{n-1}] \in W_0$ is closed.  If this
closed element is exact, i.e.,~it equals $[K,h_n]$ for some $h_n
\in W_{-1}$, we complete the $n$-th step of induction.
In general we have
\begin{equation}\label{110320:eq1}
[H,h_{n-1}]=[K,h_n]+a_n\,,
\end{equation}
where $a_n$ is a representative of the corresponding cohomology class.
Looking at \eqref{110320:eq1} more carefully,
one often can prove that one can take $a_n=0$,
so the Lenard-Magri scheme still works.

The cohomological approach to the Lenard-Magri scheme was
proposed in \cite{Kra} and \cite{Ol1}.  
A powerful machinery to compute
this cohomology and several examples can be found in \cite{DSK13b,BDSK20}.

\section{Non-local PVA}
\label{sec:7}
Non-local Poisson vertex algebras were introduced in \cite{DSK13}.
Recall the following standard notation:
for a vector space $V$,
we let $V[\lambda]$, $V[[\lambda^{-1}]]$, and $V((\lambda^{-1}))$,
be, respectively,
the spaces of polynomials in $\lambda$,
of formal power series in $\lambda^{-1}$,
and of formal Laurent series in $\lambda^{-1}$, 
with coefficients in $V$.
Furthermore,
we shall use the following notation from \cite[Sec. 2]{DSK13}:
$$
V_{\lambda,\mu}:=V[[\lambda^{-1},\mu^{-1},(\lambda+\mu)^{-1}]][\lambda,\mu]\,,
$$
namely, it is the quotient of the $\mb F[\lambda,\mu,\nu]$-module
$V[[\lambda^{-1},\mu^{-1},\nu^{-1}]][\lambda,\mu,\nu]$
by the submodule 
$(\nu-\lambda-\mu)V[[\lambda^{-1},\mu^{-1},\nu^{-1}]][\lambda,\mu,\nu]$.
%
%
Recall that we have the natural embedding 
$\iota_{\mu,\lambda}:\,V_{\lambda,\mu}\hookrightarrow V((\lambda^{-1}))((\mu^{-1}))$
defined by expanding the negative powers of $\nu=\lambda+\mu$
by geometric series in the domain $|\mu|>|\lambda|$.

Let $\mc V$ be a differential algebra. A \emph{non-local} $\lambda$-\emph{bracket} on $\mc V$ is a linear map
$\{\cdot\,_\lambda\,\cdot\}:\,\mc V\otimes \mc V\to \mc V((\lambda^{-1}))$
satisfying sesquilinearity axioms \eqref{sesqui} and the left and right Leibniz rules \eqref{lleibniz}-\eqref{rleibniz}.
Here and further an expression $\{a_{\lambda+\partial}b\}_\to c$ is interpreted as follows:
if $\{a_{\lambda}b\}=\sum_{n=-\infty}^Nc_n\lambda^n$, 
then $\{a_{\lambda+\partial}b\}_\to c=\sum_{n=-\infty}^Nc_n(\lambda+\partial)^nc$,
where we expand $(\lambda+\partial)^n$ in non-negative powers of $\partial$.
The non-local $\lambda$-bracket $\{\cdot\,_\lambda\,\cdot\}$ 
is called skewsymmetric if the skewsymmetry condition \eqref{skewsim} holds.
The RHS of the skewsymmetry condition \eqref{skewsim} should be interpreted as follows:
we move $-\lambda-\partial$ to the left and
we expand its powers in non-negative powers of $\partial$,
acting on the coefficients on the $\lambda$-bracket.
The non-local $\lambda$-bracket $\{\cdot\,_\lambda\,\cdot\}$ 
is called \emph{admissible} if
\begin{equation}\label{20110921:eq4}
\{a_\lambda\{b_\mu c\}\}\in\mc V_{\lambda,\mu}
\qquad\forall a,b,c\in\mc V\,.
\end{equation}
Here we are identifying the space $\mc V_{\lambda,\mu}$
with its image in $\mc V((\lambda^{-1}))((\mu^{-1}))$ via the embedding $\iota_{\mu,\lambda}$.
Note that, if $\{\cdot\,_\lambda\,\cdot\}$ is a skewsymmetric admissible 
non-local $\lambda$-bracket on $\mc V$,
then we also have
$\{b_\mu\{a_\lambda c\}\}\in\mc V_{\lambda,\mu}$ 
and $\{\{a_\lambda b\}_{\lambda+\mu} c\}\in\mc V_{\lambda,\mu}$,
for all $a,b,c\in\mc V$ (see \cite[Rem.3.3]{DSK13}).
A \emph{non-local PVA} is a differential algebra $\mc V$
endowed with a non-local $\lambda$-bracket,
$\{\cdot\,_\lambda\,\cdot\}:\,\mc V\otimes \mc V\to \mc V((\lambda^{-1}))$
satisfying
skewsymmetry \eqref{skewsim},
admissibility \eqref{20110921:eq4},
and Jacobi identity \eqref{jacobi}
(which is understood in the space $\mc V_{\lambda,\mu}$).

Similarly to the local case, a non-local $\lambda$-bracket on an algebra 
of differential functions $\mc V$ in the variables $\{u_i\}_{i\in I}$ (which we assume to be a domain, and we denote by $\mc K$
its field of fractions) is uniquely determined by the
$\lambda$-brackets $\{ u_{i \lambda} u_j \}$, $ i,j \in I$,
due to sesquilinearity and the left and right Leibniz rules.
The Master formula \eqref{masterformula} still gives
an explicit formula for the
$\lambda$-bracket of any $f,g \in \mc V$ in terms of $\lambda$-brackets 
of differential variables $\{u_i{}_\lambda u_j\}=H_{ji}(\lambda),\,i,j\in I$.
Furthermore, as shown in \cite[Sec. 3]{DSK13}, if the matrix pseudodifferential operator $H(\partial)=\left(H_{ij}(\partial)\right)\in\Mat_{\ell\times\ell}\mc V((\partial^{-1}))$ is \emph{rational}, namely
$$
H(\partial)=A(\partial)B(\partial)^{-1}\,,
$$
for matrix differential operators $A(\partial),B(\partial)\in\Mat_{\ell\times\ell}\mc V[\partial]$,
where $B(\partial)$ is non-degenerate, 
then the non-local $\lambda$-bracket given by the Master Formula \eqref{masterformula} is
admissible, and the skeswsymmetry axiom \eqref{skewsim} holds if $H(\partial)$ is skewadjoint, that is
$H(\partial)=-H^*(\partial)$.
%

It is shown in \cite[Sec. 4]{DSK13} that, if $H(\partial)$ is a rational skewadjoint matrix pseudodifferential operator,
then the non-local $\lambda$-bracket given by \eqref{masterformula}
defines a non-local PVA structure on $\mc V$ if and only the Jacobi
identity for each triple $u_i,u_j,u_k$ holds. The matrix pseudodifferential operator $H(\partial)$ is called a
\emph{non-local Poisson structure}.

Let $H\in\Mat_{\ell\times\ell}\mc V(\partial)$ be a non-local Poisson structure.
We say that $\xi,P\in\mc V^\ell$ are $H$-\emph{associated},
and we denote it by
$$
\xi\assr{H}P\,,
$$
if there exist
a fractional decomposition $H=AB^{-1}$ 
with $A,B\in\Mat_{\ell\times\ell}\mc V[\partial]$ 
and $B$ non-degenerate,
and an element $F\in\mc K^\ell$ such that
$\xi=BF,\,P=AF$.
An evolution equation on the variables $u=\big(u_i\big)_{i\in I}$,
\begin{equation}\label{20120124:eq5}
\frac{du}{dt}
=P\,,
\end{equation}
is called \emph{Hamiltonian} with respect to the non-local Poisson structure $H$
and the Hamiltonian functional $\tint h\in\mc V/\partial\mc V$
if $\frac{\delta h}{\delta u}\assr{H}P$.

Equation \eqref{20120124:eq5} is called \emph{bi-Hamiltonian}
if there are two compatible non-local Poisson structures $H_0$ and $H_1$,
and two local functionals $\tint h_0,\tint h_1\in\mc V/\partial\mc V$,
such that
\begin{equation}\label{20140117:eq1}
\frac{\delta h_0}{\delta u}\assr{H_1}P
\assl{H_0}
\frac{\delta h_1}{\delta u}
\,.
\end{equation}

By the chain rule, any element $f\in\mc V$ evolves according to the equation
$$
\frac{df}{dt}=\sum_{i\in I}\sum_{n\in\mb Z_{\geq0}}(\partial^nP_i)\frac{\partial f}{\partial u_i^{(n)}}
=D_f(\partial)P\,,
$$
which defines an evolutionary vector field $X_P\in\mf g_0$ via $f\mapsto X_P(f)=D_f(\partial)P$, and, integrating by parts,
a local functional $\tint f\in\mc V/\partial\mc V$
evolves according to
$$
\frac{d\tint f}{dt}=\int P\cdot\frac{\delta f}{\delta u}
\quad\bigg(=\big(P\big|\frac{\delta f}{\delta u}\big)\bigg)\,.
$$
(In the above formula $(\cdot|\cdot)$ denotes the non-degenerate pairing in \eqref{pairing}.)
An \emph{integral of motion} for the Hamiltonian equation \eqref{20120124:eq5}
is a local functional $\tint f\in\mc V/\partial\mc V$
which is constant in time, i.e. such that $(P|\frac{\delta f}{\delta u})=0$.
The usual requirement for \emph{integrability}
is to have
sequences $\{\tint h_n\}_{n\in\mb Z_{\geq0}}\subset\mc V/\partial\mc V$ 
and $\{P_n\}_{n\in\mb Z_{\geq0}}\subset\mc V^\ell$,
starting with $\tint h_0=\tint h$ and $P_0=P$,
such that
\begin{enumerate}[(C1)]
\item
$\frac{\delta h_n}{\delta u}\assr{H}P_n$ for every $n\in\mb Z_{\geq0}$,
\item
$[X_{P_m},X_{P_n}]=0$ for all $m,n\in\mb Z_{\geq0}$, where $[\,\cdot,\cdot\,]$ is the commutator of vector fields defined in \eqref{commvec},
\item
$(P_m\,|\,\frac{\delta h_n}{\delta u})=0$ for all $m,n\in\mb Z_{\geq0}$.
\item
The elements $P_n$ span an infinite dimensional subspace of $\mc V^\ell$.
\end{enumerate}
In this case, we have an \emph{integrable hierarchy} of Hamiltonian equations
$$
\frac{du}{dt_n} = P_n\,,\,\,n\in\mb Z_{\geq0}\,.
$$
Elements $\tint h_n$'s are called \emph{higher Hamiltonians},
the $P_n$'s are called \emph{higher symmetries},
and the condition $(P_m\,|\,\frac{\delta h_n}{\delta u})=0$
says that $\tint h_m$ and $\tint h_n$ are \emph{in involution}.
Note that (C4) implies that elements $\frac{\delta h_n}{\delta u}$
span an infinite dimensional subspace of $\mc V^\ell$.
The converse holds provided that either $H_0$ or $H_1$ is non-degenerate.

Suppose we have a bi-Hamiltonian equation \eqref{20120124:eq5},
associated to the compatible Poisson structures $H_0,H_1$
and the Hamiltonian functionals $\tint h_0,\tint h_1$,
in the sense of equation \eqref{20140117:eq1}.
The \emph{Lenard-Magri scheme of integrability}
consists in finding
sequences $\{\tint h_n\}_{n\in\mb Z_{\geq0}}\subset\mc V/\partial\mc V$ 
and $\{P_n\}_{n\in\mb Z_{\geq0}}\subset\mc V^\ell$,
starting with $P_0=P$ and the given Hamiltonian functionals $\tint h_0,\tint h_1$,
satisfying the following recursive relations:
\begin{equation}\label{20130604:eq7}
\frac{\delta h_{n-1}}{\delta u}\assr{H_1}P_n
\assl{H_0}\frac{\delta h_n}{\delta u}
\,\,\,\,
\text{ for all } n\in\mb Z_{\geq0}
\,.
\end{equation}
In this case,
we have the corresponding bi-Hamiltonian hierarchy
\begin{equation}\label{20130604:eq6}
\frac{du}{dt_n}=P_n\,\in\mc V^\ell
\,\,,\,\,\,\,
n\in\mb Z_{\geq0}
\,,
\end{equation}
all Hamiltonian functionals $\tint h_n,\,n\geq-1$,
are integrals of motion for all equations of the hierarchy,
and they are in involution with respect to both Poisson structures $H_0$ and $H_1$,
and all commutators $[P_m,P_n]$ are zero, provided that $\Ker B^*\cap\Ker D^*=0$, where
$H_0=A(\partial)B(\partial)^{-1}$ and $H_1=C(\partial)D(\partial)^{-1}$ (see \cite[Sec. 7]{DSK13}).
Hence, in this situation \eqref{20130604:eq6} is an integrable hierarchy
of compatible evolution equations,
provided that condition (C4) holds.

As an example, we describe in detail how to construct the non-linear Schroedinger (NLS) hierarchy.
Let $\mc V$ be the algebra of differential polynomials in two variables $u$ and $v$. A compatible pair of non-local Poisson structures is ($\alpha,\beta\in\mb F$):
\begin{equation}\label{20121109:eq1}
\begin{array}{l}
\displaystyle{
H_0=
\partial\id
+2\alpha\left(\begin{array}{cc} 
v\partial^{-1}\circ v & -v\partial^{-1}\circ u \\
-u\partial^{-1}\circ v & u\partial^{-1}\circ u
\end{array}\right)
\,}\\
\displaystyle{
H_1=
\left(\begin{array}{cc} 0 & -1 \\ 1 & 0 \end{array}\right)
+2\beta\left(\begin{array}{cc} 
v\partial^{-1}\circ v & -v\partial^{-1}\circ u \\
-u\partial^{-1}\circ v & u\partial^{-1}\circ u
\end{array}\right)
\,.}
\end{array}
\end{equation}
We have the following $H_0$ and $H_1$-associations:
$0\assr{H_0}P_0\assl{H_1}\frac{\delta h_0}{\delta u}\assr{H_0}P_1\assl{H_1}\frac{\delta h_1}{\delta u}\assr{H_0}P_2$,
where 
\begin{align*}
\begin{array}{l}
\displaystyle{
P_0=2\alpha^2\left(\begin{array}{c} -v \\ u \end{array}\right)
\,\,,\,\,\,\,
\tint h_0=\frac{1}{2}\tint (u^2+v^2)
\,,} \\
\displaystyle{
P_1= \left(\begin{array}{c} u' \\ v'\end{array}\right)
\,\,,\,\,\,\,
\tint h_1=
\int\Big(
uv'+\frac{\beta}{4}(u^2+v^2)^2
\Big)
\,,} \\
\displaystyle{
P_2=
\left(\begin{array}{c} 
v''+\alpha v(u^2+v^2)+\beta\big(u(u^2+v^2)\big)^\prime  \\ 
-u''-\alpha u(u^2+v^2)+\beta\big(v(u^2+v^2)\big)^\prime 
\end{array}\right)
\,.}
\end{array}
\end{align*}
It is shown in \cite[Sec. 7]{DSK13} that we can include $\tint h_0,\tint h_1$ and $P_0,P_1,P_2$ in infinite
sequences $\{\tint h_n\}_{n\in\mb Z_{\geq0}}\subset\mc V/\partial\mc V$ 
and $\{P_n\}_{n\in\mb Z_{\geq0}}\subset\mc V^\ell$ such that the above conditions (C1)-(C4) hold.
We thus get an integrable hierarchy of bi-Hamiltonian PDE \eqref{20130604:eq6}, whose first non-trivial
equation is
\begin{equation}\label{20121109:eq5}
\frac{d}{dt_2}\left(
\begin{array}{c}
u\\
v
\end{array}
\right)=
\left(\begin{array}{c} 
v''+\alpha v(u^2+v^2)+\beta\big(u(u^2+v^2)\big)^\prime  \\ 
-u''-\alpha u(u^2+v^2)+\beta\big(v(u^2+v^2)\big)^\prime 
\end{array}\right)
\,.
\end{equation}
If we view $u$ and $v$ as real valued functions, and we consider the complex valued function
$\psi=u+iv$, the system \eqref{20121109:eq5} can be written as the following PDE:
$$
i\frac{d\psi}{dt}=\psi''+\alpha\psi|\psi|^2+i\beta(\psi|\psi|^2)^\prime\,,
$$
which, for $\beta=0$, is the NLS equation
(see e.g. \cite{TF86,Dor93,BDSK09}).
The case $\beta\neq0$ (derivative NLS) has been studied by many authors as well, 
the first reference that we know is apparently \cite{KN78}.

\section{Dirac reduction of non-local PVA}
\label{sec:8}

Let $H(\partial)\in\Mat_{\ell\times\ell}\mc V(\partial)$
be a Poisson structure on $\mc V$.
Let $\{\cdot\,_\lambda\,\cdot\}_H$ be the corresponding PVA $\lambda$-bracket on $\mc V$
given by the Master Formula \eqref{masterformula}.
Let $\theta_1,\dots,\theta_m$ be some elements of $\mc V$,
and let $\mc I=\langle\theta_1,\dots,\theta_m\rangle_{\mc V}\subset\mc V$ 
be the differential ideal generated by them.
Consider the following rational matrix pseudodifferential operator
\begin{equation}\label{20130529:eq1}
C(\partial)=D_\theta(\partial)\circ H(\partial)\circ D_\theta^*(\partial)\,
\in\Mat_{m\times m}\mc V(\partial)\,,
\end{equation}
where $D_\theta(\partial)$ is the $m\times\ell$ matrix differential operator
of Frechet derivatives of the elements $\theta_i$'s defined by \eqref{20111020:eq1} and $D_\theta^*(\partial)\in\Mat_{\ell\times m}\mc V[\partial]$ is its adjoint.
Recalling the Master Formula \eqref{masterformula},
we get that $C(\partial)$ has matrix elements with symbols
$$
C_{\alpha\beta}(\lambda)=\{\theta_{\beta}{}_{\lambda}\theta_{\alpha}\}_H\,.
$$
Note also that, by the skewadjointness of $H$,
the corresponding $\lambda$-bracket $\{\cdot\,_\lambda\,\cdot\}_H$
is skewsymmetric, hence $C(\partial)$ is a skewadjoint pseudodifferential operator.

We shall assume that the matrix $C(\partial)$ in \eqref{20130529:eq1}
is invertible in $\Mat_{m\times m}\mc V((\partial^{-1}))$,
and we denote its inverse by
$C^{-1}(\partial)=\big((C^{-1})_{\alpha\beta}(\partial)\big)_{\alpha,\beta=1}^m
\in\Mat_{m\times m}\mc V((\partial^{-1}))$.
The \emph{Dirac modification} of the Poisson structure $H\in\Mat_{\ell\times\ell}\mc V(\partial)$
by the \emph{constraints} $\theta_1,\dots,\theta_m$
is the following skewadjoint $\ell\times\ell$ matrix pseudodifferential operator:
\begin{equation}\label{20130529:eq3}
H^D(\partial)
=
H(\partial)+B(\partial)\circ C^{-1}(\partial)\circ B^*(\partial)\,,
\end{equation}
where $B(\partial)=H(\partial)\circ D_\theta^*(\partial)
\in\Mat_{\ell\times m}\mc V(\partial)$.
The matrix pseudodifferential operator $H^D(\partial)$ is skewadjoint and rational.
The corresponding $\lambda$-bracket, given by the Master Formula \eqref{masterformula}, 
is
\begin{equation}\label{dirac}
\{f_{\lambda}g\}_H^D
=\{f_{\lambda}g\}_H
-\sum_{\alpha,\beta=1}^m
{\{{\theta_{\beta}}_{\lambda+\partial}g\}_H}_{\to}
(C^{-1})_{\beta\alpha}(\lambda+\partial)
{\{f_{\lambda}\theta_{\alpha}\}_H}\,.
\end{equation}
It is shown in \cite[Sec. 2]{DSKV13c} that
the Dirac modified $\lambda$-bracket \eqref{dirac} satisfies the Jacobi identity \eqref{jacobi},
thus the Dirac modification $H^D(\partial)$ is a non-local Poisson structure on $\mc V$.
Furthermore, all the elements $\theta_i,\,i=1,\dots,m$, are central 
with respect to the Dirac modified $\lambda$-bracket,
and the differential ideal $\mc I=\langle\theta_1,\dots,\theta_m\rangle_{\mc V}\subset\mc V$,
generated by $\theta_1,\dots,\theta_m$,
is an ideal with respect to the Dirac modified $\lambda$-bracket $\{\cdot\,_\lambda\,\cdot\}^D$.
Hence, the quotient space $\mc V/\mc I$ is a PVA,
with $\lambda$-bracket induced by $\{\cdot\,_\lambda\,\cdot\}^D$,
called the \emph{Dirac reduction} of $\mc V$ 
by the constraints $\theta_1,\dots,\theta_m$.
This construction is a PVA analogue of the classical Dirac reduction of Poisson algebras \cite{Dirac}.

Let now $(H_0,H_1)$ be a bi-Poisson structure on $\mc V$
and let $\theta_1,\dots,\theta_m\in\mc V$ be central elements for $H_0$.
Suppose that the matrix pseudodifferential operator (cf. \eqref{20130529:eq1})
$C(\partial)=D_\theta(\partial)\circ H_1(\partial)\circ D_\theta^*(\partial)$
is invertible.
Then we can consider the Dirac modified Poisson structure $H_1^D$ (cf. \eqref{20130529:eq3}),
and the corresponding $\lambda$-bracket $\{\cdot\,_\lambda\,\cdot\}_1^D$ (cf. \eqref{dirac}).
It is proved in \cite[Sec. 7]{DSKV13c} that the matrices $H_0$ and $H_1^D$ form a compatible pair of Poisson structures on $\mc V$
and that the differential algebra ideal
$\mc I=\langle\theta_1,\dots,\theta_m\rangle_{\mc V}$
is a PVA ideal for both the $\lambda$-brackets 
$\{\cdot\,_\lambda\,\cdot\}_0$
and $\{\cdot\,_\lambda\,\cdot\}_1^D$,
hence we have the induced compatible PVA $\lambda$-brackets on $\mc V/\mc I$.

Let $(H_0,H_1)$ be a \emph{local} bi-Poisson structure
(i.e. consisting of matrix differential operators).
Suppose that we have a bi-Hamiltonian hierarchy 
$\frac{du}{dt_n}=P_n\in\mc V^\ell$, $n\in\mb Z_{\geq0}$,
with respect to $(H_0,H_1)$,
and let $\tint h_n\in\mc V/\partial\mc V$
be a sequence of integrals of motion
satisfying the Lenard-Magri recursive condition \eqref{20130604:eq7}.
Let $\theta_1,\dots,\theta_m\in\mc V$ be central elements for $H_0$.
Assume that the matrix
$C(\partial)=D_\theta(\partial)\circ H_1(\partial)\circ D_\theta^*(\partial)\in\Mat_{m\times m}\mc V[\partial]$
is invertible in $\Mat_{m\times m}\mc V((\partial^{-1}))$.
Then,
$H_1^D=H_1+B(\partial)C^{-1}(\partial)B^*(\partial)$,
where $B(\partial)=H_1(\partial)\circ D_\theta^*(\partial)$,
is a non-local Poisson structure on $\mc V$ compatible to $H_0$.
A sufficient condition to have the (Dirac reduced) Lenard-Magri recursion
$$
\frac{\delta h_{n-1}}{\delta u}\assr{H_1^D}P_n
\assl{H_0}
\frac{\delta h_n}{\delta u}
\,\,\,\,
\text{ for all } n\in\mb Z_{\geq0}
\,,
$$
is that $\Ker B(\partial)$ and $\Ker C(\partial)$
have zero intersection over the linear closure $\widetilde{\mc K}$ of $\mc K$, see \cite[Prop. 2.5]{DSKV14}.

As an example, we explain how to get integrability of the NLS hierarchy from the integrability of the homogeneous
Drinfeld-Sokolov hierarchy associated to the Lie algebra $\mf{sl}_2$, see \cite{DS85,DSKV13a} and Section \ref{sec:12}.
Let $\mc V=\mc V(\mf{sl}_2)$ be the
affine PVA associated to the Lie algebra $\mf{sl}_2$, with its trace form $(a|b)=\Tr(ab)$, and the element $s=\frac{h}{2}$
described in Section \ref{sec:2}. It is the algebra of differential polynomials in the variables $h,e,f$, the standard basis of $\mf{sl}_2$, endowed with the compatible pair of local Poisson structures
\begin{align*}
\begin{array}{l}
\displaystyle{
H_0=
\left(\begin{array}{ccc} 
2\partial & -2 e & 2f\\
2e & 0& \partial -h\\
-2 f& \partial +h&0
\end{array}\right)
\,,}
\quad\displaystyle{
H_1=
\left(\begin{array}{ccc}
0&0&0
\\
0& 0 & -1
\\ 0&1& 0 \end{array}\right)
\,.}
\end{array}
\end{align*}
Note that these Poisson structures correspond to the $\lambda$-bracket \eqref{lambda_affine}.
Let $\mc I=\langle h\rangle_{\mc V}$ be the differential ideal generated by $h$. Then $\overline{\mc V}=\mc V/\mc I$
is the algebra of differential polynomials in the variables $e$ and $f$. Note that
$$
C(\lambda)=\{h_\lambda h\}_{H_1}=2\lambda
\,,
$$
is the symbol of an invertible pseudodifferential operator, while $h$ is central with respect to the Poisson structure $H_0$. Hence, we get Dirac reduced compatible Poisson structures on $\overline{\mc V}$. By \eqref{20130529:eq3},
they  are 
\begin{align*}
\begin{array}{l}
\displaystyle{
H_0^D=
\left(\begin{array}{cc} 
2e\partial^{-1}\circ e &\partial -2e\partial^{-1} \circ f\\
\partial -2f\partial^{-1} \circ e& 2f\partial^{-1}\circ f
\end{array}\right)
\,,}
\quad\displaystyle{
H_1=
\left(\begin{array}{cc}
 0 & -1\\
 1 &0\end{array}\right)
\,.}
\end{array}
\end{align*}
After the change of variables $u=\frac{e+f}{\sqrt 2}$, $v=\frac{e-f}{\sqrt 2}$, these become the Poisson structures \eqref{20121109:eq1} of the NLS equation \eqref{20121109:eq5}. In fact, as proven in \cite{DSKV14}, the
NLS hierarchy can be obtained as a Dirac reduction of an integrable hierarchy associated to $\mc V(\mf{sl}_2)$.

\section{Adler identity and integrability}
\label{sec:9b}

In the paper \cite{DSKV16}  a general method of constructing
integrable (bi)Hamiltonian hierarchies of PDE's was proposed.
It combines two most famous approaches.
The first one is the Gelfand-Dickey technique,
based on the Lax pair method,
which consists in taking fractional powers of pseudodifferential operators \cite{GD76,Dic03},
and the second one is the classical Hamiltonian reduction technique, 
combined with the Zakharov-Shabat method,
as developed by Drinfeld and Sokolov \cite{DS85}.

The central notion of the method is that of a
matrix pseudodifferential operator of \emph{Adler type},
see \cite[Sec. 2]{DSKV15}.
It has been derived there starting from Adler's formula
for the second Poisson structure for the $N$-th KdV hierarchy \cite{Adl79}.
A pseudodifferential operator $A(\partial)$ over a differential algebra $\mc V$ 
is called of \emph{Adler type},
with respect to a $\lambda$-bracket $\{\cdot\,_\lambda\,\cdot\}$ on $\mc V$, if
\begin{equation}\label{eq:adler-scalar}
\begin{split}
\{A(z)_\lambda A(w)\}
& = A(w+\lambda+\partial)\iota_z(z\!-\!w\!-\!\lambda\!-\!\partial)^{-1}A^*(\lambda-z)
\\
& - A(z)\iota_z(z\!-\!w\!-\!\lambda\!-\!\partial)^{-1}A(w)
\,.
\end{split}
\end{equation}
Here $A(z)$ is the symbol of $A(\partial)$,
$A^*(\partial)$ is the formal adjoint of $A(\partial)$,
and $\iota_z$ denotes the expansion in the geometric series for large $z$.

The definition of a matrix pseudodifferential operator of Adler type
is similar, see \cite[Sec. 4]{DSKV15}. In \cite{DSKV18} the definition has been generalized to self/skew-adjoint matrix pseudodifferential
operators.

The first basic property of an Adler type matrix pseudodifferential operator $A(\partial)$,
is that the $\lambda$-bracket on $\mc V$
restricted to the subalgebra $\mc V_1$
generated by the coefficients of the entries of $A(\partial)$
satisfies the skewcommutativity \eqref{skewsim} and Jacobi identity \eqref{jacobi} axioms of a PVA.
Thus an Adler type operator automatically provides $\mc V_1$
with a Poisson structure.

For example, it is noted in \cite{DSKV15}
that, for each positive integer $N$, the ``generic''
pseudodifferential operator
(resp. ``generic'' differential operator)
$$
L_N(\partial)
=
\sum_{j=-\infty}^N u_j\partial^j
\qquad\Big(\text{resp. }\,\, 
L_{(N)}(\partial)
=
\sum_{j=0}^N u_j\partial^j
\,\Big)
\,,\qquad\text{ with }
u_N=1
\,,
$$
is of Adler type.
Consequently one automatically gets on the algebra of differential polynomials
in $\{u_j\,|\,-\infty<j<N\}$
(resp. $\{u_j\,|\,0\leq j<N\}$)
the so called $2$-nd Poisson structure for the KP hierarchy
(resp. for the $N$-th KdV hierarchy).
This Poisson structure for the $N$-th KdV hierarchy has been conjectured by Adler
in \cite{Adl79},
and subsequently proved by a lengthy calculation in \cite{GD78}.
The second Poisson structures for the KP hierarchy
were discovered by Radul in \cite{Rad87}.
Note that these structures are different for different $N$,
though the corresponding hierarchies of equations are 
the same up to a change of variables.
Note also that the $1$-st Poisson structure for the KP hierarchy
was previously discovered in \cite{Wat83},
and it is well known to be easily derived from the $2$-nd structure.
In fact, this is the case in general:
if $A(\partial)+\epsilon$ is of Adler type for any constant $\epsilon$,
then we call $A(\partial)$ of bi-Adler type,
and we have on $\mc V_1$ a bi-PVA structure.

The second basic property of an Adler type (matrix) pseudodifferential
operator $A(\partial)$
is that it provides a hierarchy of compatible
Lax type equations via the method of fractional powers:
\begin{equation}\label{eq:intro2}
\frac{dA(\partial)}{dt_{n,k}}
=
\big[\big(A(\partial)^{\frac nk}\big)_+,A(\partial)\big]
\,,
\end{equation}
where $k,n$ are positive integers,
see \cite[Sec. 3]{DSKV15}.
Here, as usual, the subscript $+$ stands for the differential part
of a pseudodifferential operator, and $A(\partial)^{\frac nk}$ is a (matrix) pseudodifferential operator such that $\left(A(\partial)^{\frac nk}\right)^k=A(\partial)^{n}$.
For example, for $A(\partial)=L_1(\partial)$ and $k=1$
(resp. $A(\partial)=L_{(N)}(\partial)$ and $k=N$),
equation \eqref{eq:intro2} is Sato's KP hierarchy \cite{Sat81}
(resp. Gelfand-Dickey's $N$-th KdV hierarchies \cite{GD76,Dic03}).
The name reflects the fact that for $N=2$ one gets the KdV hierarchy discussed in Section \ref{sec:4}.

Furthermore, equations \eqref{eq:intro2}
are Hamiltonian with respect to the Poisson structure described above,
and they are bi-Hamiltonian if $A(\partial)$ is of bi-Adler type.

The third basic property of an Adler type pseudodifferential operator $A(\partial)$
is that it provides an infinite set of conserved densities
for the hierarchy \eqref{eq:intro2}:
\begin{equation}\label{eq:intro3}
h_{n,k}
=
\frac{-k}{n}\Res_\partial\Tr \big(A(\partial)^{\frac nk}\big)
\,,\,\,n,k\in\mb Z_{>0}\,,
\end{equation}
where, for a matrix pseudodifferential operator $B(\partial)=\sum_{n\leq N}B_n \partial^n$, its residue is
$\Res B(\partial)=B_{-1}$.
Moreover, provided that $A(\partial)$ is bi-Adler,
these conserved densities satisfy the (generalized)
Lenard-Magri scheme \eqref{eq:LM}.

The approach just described allows one not only to recover the basic results
of the KP theory and $N$-th KdV theory,
including the matrix case,
but to go far beyond that
in constructing new integrable (bi)Hamiltonian hierarchies of Lax type equations.
Of course, for this aim one needs to construct new Adler type operators.

The basic example is the family of $N\times N$-matrix differential operators
\begin{equation}\label{eq:intro4}
A_{S}(\partial)
=
\id_N\partial+\sum_{i,j=1}^Nu_{ji}E_{ij}+S
\,\in\Mat_{N\times N}\mc V[\partial]
\,,
\end{equation}
where $\mc V=\mc V(\mf{gl}_N)$ is the algebra of differential polynomials in the basis $u_{ij}$, given by elementary matrices and $S\in\Mat_{N\times N}\mb F$.
The operator $A_S(\partial)$ is of Adler type
with respect to the affine PVA $\lambda$-bracket \eqref{lambda_affine} on $\mc V$,
where $(a|b)=\Tr(ab)$ and $s=S$ (and $z=1$).
Unfortunately, the operator $A_S(\partial)$ does not have ``good'' $K$-th roots.
The way out is provided by yet another remarkable property of a square matrix non-degenerate
Adler type operator $A(\partial)$:
its inverse is of Adler type as well, with respect to the negative of the $\lambda$-bracket
for $A(\partial)$ on $\mc V$.
It follows that the generalized quasideterminants of $A(\partial)$
are of Adler type with respect to the same $\lambda$-bracket as for $A(\partial)$, see \cite[Sec. 3]{DSKV16}.

Recall that a \emph{quasideterminant} of an invertible matrix $A$
over an associative (not necessarily commutative) unital ring
is the inverse (if it exists) of an entry of $A^{-1}$.
This notion was the key to the systematic development of linear algebra
over non-commutative associative rings
in a series of papers by Gelfand and Retakh, and their collaborators,
see \cite{GGRW05} for a review.

In the paper \cite[Sec. 4]{DSKV16} the following generalization of quasideterminants is used:
if $I\in\Mat_{N\times M}\mb F$ and $J\in\Mat_{M\times N}\mb F$
are rectangular matrices of rank $M\leq N$,
the $(I,J)$ quasideterminant of an invertible matrix $A$ is $|A|_{IJ}=(JA^{-1}I)^{-1}$,
assuming that the $M\times M$ matrix $JA^{-1}I$ is invertible.

To demonstrate the method of integrability developed in \cite{DSKV16},
consider, for example, the matrix $A_{\epsilon S}$ defined in \eqref{eq:intro4},
which is of Adler type with respect to the bi-PVA $\lambda$-bracket
$\{\cdot\,_\lambda\,\cdot\}_\epsilon=\{\cdot\,_\lambda\,\cdot\}_0+\epsilon\{\cdot\,_\lambda\,\cdot\}_\infty$
on $\mc V$, defined by \eqref{lambda_affine} with $S$ replaced by $\epsilon S$.
Let $I\in\Mat_{N\times M}\mb F$ and $J\in\Mat_{M\times N}\mb F$
be rectangular matrices of rank $M\leq N$ such that $S=IJ$
($I$ and $J$ are uniquely defined up to a change of basis in $\mb F^M$).
Then,
the generalized quasideterminant $|A|_{IJ}$ is 
a matrix pseudodifferential operator of bi-Adler type
with respect to the same bi-PVA $\lambda$-bracket $\{\cdot\,_\lambda\,\cdot\}_\epsilon$.
As a consequence, 
we automatically get, by the above mentioned method of fractional powers (cf. \eqref{eq:intro3}),
a sequence of Hamiltonian functionals
$$
\tint h_n=-\frac1n\tint\Res_\partial\Tr (|A(\partial)|_{IJ})^n\,,\quad n\geq1\,,\quad \tint h_0=0\,,
$$
which are in involution with respect to both PVA $\lambda$-brackets
$\{\cdot\,_\lambda\,\cdot\}_0$ and $\{\cdot\,_\lambda\,\cdot\}_\infty$,
they satisfy the Lenard-Magri recursion relation \eqref{eq:LM}
in the subalgebra $\mc V_1\subset\mc V$ generated by the coefficients of $|A|_{IJ}$,
and therefore produce an integrable hierarchy of bi-Hamiltonian equations
$$
\frac{du}{dt_n}
=
\{{\tint h_n}_\lambda u\}_0
=
\{{\tint h_{n+1}}_\lambda u\}_\infty\,,
\quad
n\geq0\,,u\in\mc V_1
\,.
$$
In \cite{DSKV16a,DSKV18}, this method was applied to construct an integrable hierarchy of bi-Hamiltonian equations
associated to any classical affine $\mc W$-algebra (introduced in Section \ref{sec:11})
associated to a classical Lie algebra $\mf g$ and
any nilpotent element $f\in\mf g$.
The Adler identity \eqref{eq:adler-scalar} has revealed useful in constructing $\tau$-functions for these
hierarchies in the case of $\mf g=\mf{gl}_N$, see \cite{tau}.

\section{Classical affine \texorpdfstring{$\mc W$}{W}-algebras}
\label{sec:11}
Let $\mf g$ be a finite dimensional Lie algebra with a non-degenerate symmetric invariant bilinear form
$(\cdot\,|\,\cdot)$.
Let $\mf s=\{e,h,f\}\subset\mf g$ be an $\mf{sl}_2$-triple.
We have the $\ad \frac h2$-eigenspace decomposition
\begin{equation}\label{eq:dec}
\mf g=\bigoplus_{k\in \frac{1}{2}\mb Z}\mf g_{k}
\,,
\qquad
\mf g_k=\big\{a\in\mf g\mid [h,a]=2ka\big\}
\,.
\end{equation}
The largest $d\in\frac{1}{2}\mb Z$ such that $\mf g_d\neq0$ is called the \emph{depth}.

Let $\omega$ be
the skew-symmetric non-degenerate bilinear form on $\mf g_{\frac12}$
given by $\omega(a,b)=(f|[a,b])$.
Fix an isotropic (with respect to $\omega$) subspace $\mf l\subset\mf g_{\frac12}$
and denote by $\mf l^\perp=\{a\in\mf g_{\frac12}\mid \omega(a,b)=0 \text{ for all }b\in\mf l\}\subset\mf g_{\frac12}$
its orthogonal complement with respect to $\omega$.
Consider the following nilpotent subalgebras of $\mf g$:
$$
\mf m = \mf l \oplus \mf g_{\geq1}\subset\mf n = \mf l^\perp\oplus\mf g_{\geq1}\,,
$$
where $\mf g_{\geq1}=\oplus_{k\geq1}\mf g_{k}$.

Fix an element $E\in\mf z(\mf n)$ (the centralizer of $\mf n$ in $\mf g$)
and consider the affine bi-PVA $\mc V(\mf g)$ 
from Section \ref{sec:2} with $s=E$.
Let also $I$ be the differential algebra ideal generated by the set
$$
\big\langle m-(f|m) \big\rangle_{m\in\mf m}\,\subset\mc V(\mf g)
\,.
$$
Consider the space
$$
\widetilde{\mc W}
=
\big\{w\in\mc V(\mf g)\,\big|\, \{a_\lambda w\}_0\in I[\lambda]\,,\text{ for every }a\in\mf n\big\}
\,\subset\,
\mc V(\mf g)
\,.
$$
It is shown in \cite[Sec. 3]{DSKV13a}
that
$\widetilde{\mc W}\subset\mc V(\mf g)$ is a bi-PVA subalgebra of $\mc V(\mf g)$
and
$I\subset\widetilde{\mc W}$ is a bi-PVA ideal.
The corresponding \emph{classical affine} $\mc W$-\emph{algebra}
is defined as the quotient
$$
\mc W
:=\mc W(\mf g,f)=
\widetilde{\mc W}/I
\,,
$$
which has a natural structure of a bi-PVA,
with the induced PVA $\lambda$-bracket $\{\cdot\,_\lambda\,\cdot\}_z$
from \eqref{lambda_affine}.

As a differential algebra, $\mc W$ is the algebra of differential polynomials in $\ell$ variables
$w_1,\dots,w_\ell$, where $\ell$ is the dimension of $\mf g^f$, the centralizer of $f$ in $\mf g$.
In \cite{structure}, explicit formulas for the $\lambda$-brackets $\{{w_i}_\lambda w_j\}_z$, $1\leq i,j\leq\ell$, are found.

By the Jacobson-Morozov Theorem 
any non-zero nilpotent element $f$ is part of an
$\mf{sl}_2$-triple and by Kostant's Theorem 
all $\mf{sl}_2$-triples containing $f$ are conjugate by the centralizer of $f$ in $G$,
the adjoint group of $\mf g$. It follows that all the above constructions depend only on the $G$-orbit of $f$,
and not on the chosen $\mf{sl}_2$-triple.

For example, the classical affine $\mc W$-algebra associated to the Lie algebra $\mf{sl}_2$, with its trace form
$(a|b)=\Tr(ab)$, its nilpotent element $f$ and $E=\frac e2$ is the algebra of differential polynomials in one variable $u$ endowed
with the bi-PVA structure given by $H_0(\partial)=u'+2u\partial-\frac12\partial^3$, the Virasoro-Magri Poisson structre 
\eqref{intro:eq7} for $c=-\frac12$, and $H_\infty(\partial)=\partial$, the GFZ Poisson structure.

\section{Integrable triples and integrability of classical affine \texorpdfstring{$\mc W$}{W}-algebras}
\label{sec:12}

In the seminal paper \cite{DS85}, Drinfeld and Sokolov constructed integrable hierarchies of bi-Hamiltonian PDE,
nowadays known as \emph{Drinfeld-Sokolov (DS) hierarchies}, for the classical
affine $\mc W$-algebra associated to a simple Lie algebra $\mf g$, its principal nilpotent element $f$ and a highest
root vector $E$. (In the case of $\mf{sl}_2$ one gets the KdV hierarchy as explained before.)

Subsequently, the construction of these hierarchies has been generalized by many authors to other classes of nilpotent elements,
see \cite{dGHM92,FHM92,BdGHM93,FGMS95,FGMS96,DSKV13a}, leading to \emph{generalized DS hierarchies}.
In this section we review the construction,  provided in \cite{mamuka1}, 
of generalized DS hierarchies for any integrable triple associated to $f$ defined below. Integrable triples are classified, up to
equivalence, in \cite{mamuka2}. They exist for all but one nilpotent conjugacy class in $G_2$, one in $F_4$ and five in $E_8$. Hence,
we have generalized DS hierarchies for all simple Lie algebras $\mf g$ and nilpotent elements $f$, except possibly for the
above mentioned nilpotent conjugacy classes.

An \emph{integrable triple} associated to $f$ is $(f_1,f_2,E)$, where $f_1,f_2\in \mf g_{-1}$ and
 $E\in\mf g_{\geq\frac12}$ is a non-zero homogeneous element, such that the following three properties hold:
\begin{enumerate}[(i)]
\item $f=f_1+f_2$ and $[f_1,f_2]=0$,
\item $[E, \mf g_{\geq 1}]=0$ and the centralizer $\mf l^\perp$ of $E$ in $\mf g_{\frac{1}{2} }$ is
coisotropic with respect to the bilinear form $\omega$.
\item $f_1+E$ is semisimple and $[f_2, E]=0$.
\end{enumerate}  
In this case $E$ is called an \emph{integrable element} for $f$.

Note that for an integrable triple $(f_1,f_2,E)$ the decomposition $f+E=(f_1+E)+ f_2$ is  a Jordan decomposition
of $f+E$, and that  $E$ is a central element of the subalgebra $\mf n:=\mf l^\perp\oplus\mf g_{\geq1}$.

To construct generalized DS hierarchies we use a generalization of the DS construction developed in \cite{DSKV13a}.
Let $\mb K = \mb F ((z^{-1}))$ be the field of formal Laurent series in $z^{-1}$ over $\mb F$. Consider the Lie algebra $\mf g ((z^{-1}))
=\mf g\otimes_{\mb F} \mb K$.
Since, $f_1+E\in\mf g$ is semisimple, 
$f_1+zE\in\mf g((z^{-1}))$ is also semisimple.
We thus have the direct sum decomposition
$$
\mf g((z^{-1}))=\mf h\oplus\mf h^\perp
\,,
$$
where
$$
\mf h
:=
\ker\ad (f_1+zE)
\,\,\text{ and }\,\,
\mf h^\perp
:=
\im\ad(f_1+zE)
\,.
$$
(The notation $\mf h^\perp$ relates to the fact that $\im\ad(f_1+zE)$
is the orthogonal complement of $\ker\ad (f_1+zE)$
with respect to the non-degenerate symmetric invariant
$\mb K$-valued bilinear form $(\cdot\,|\,\cdot)$
on $\mf g((z^{-1}))$, extending the form $(\cdot\,|\,\cdot)$ on $\mf g$ by $\mb K$-bilinearity.)

If $E\in\mf g_k$, we extend the $\frac12\mb Z$-grading \eqref{eq:dec}
to $\mf g((z^{-1}))$ by letting $z$ have degree $-k-1$,
so that $f+zE$ and $f_1+zE$ are homogeneous of degree $-1$. Note that, since $f_1+zE$ is homogeneous, we have the corresponding decompositions
of $\mf h$ and $\mf h^\perp$, and we denote by $\mf h_i$ and $\mf h^\perp_i$, $i\in\frac12\mb Z$,
 the homogeneous components of these decompositions.

The key observation in the construction is that for every element $A(z)\in\mf g((z^{-1}))$, there exist unique $h(z)\in\mf h$ and $U(z)\in\mf h^\perp$ such that
\begin{equation}\label{toprove}
h(z)+[f+zE, U(z)]=A(z)
\,.
\end{equation}
Consider the Lie algebra 
$$
\widetilde{\mf g}=\mb F\partial\ltimes\big(\mc W\otimes\mf g((z^{-1}))\big)\,,
$$
where $\mc W=\mc W(\mf g,f)$ is viewed as an abelian Lie algebra and $\partial$ acts on the first factor, namely for $f\in\mc W$ and $a(z)\in\mf g((z^{-1}))$
we have $[\partial, f\otimes a(z)]=(\partial f)\otimes a(z)$.
For every $U(z)\in \mc W\otimes\mf g((z^{-1}))_{>0}\subset\widetilde{\mf g}$,
we have a well-defined automorphism of the Lie algebra $\widetilde{\mf g}$
given by $e^{\ad U(z)}$.
We extend the bilinear form $(\cdot\,|\,\cdot)$ on $\mf g((z^{-1}))$
to a map
\begin{equation}\label{20200625:eq1}
(\cdot\,|\,\cdot)
\,:\,\,
\big(\mc W\otimes\mf g((z^{-1}))\big)
\times
\big(\mc W\otimes\mf g((z^{-1}))\big)
\to
\mc W((z^{-1}))
\,,
\end{equation}
given by 
$(g\otimes a(z)|h\otimes b(z))
=
gh(a(z)|b(z))$.

Let $\{q_i\}_{i=1,\dots,\ell}$ be a basis of the centralizer of $f$, and let
$\{q^i\}_{i=1,\dots\ell}$ be the dual (with respect to $(\cdot\,|\,\cdot)$) basis of $\mf z(e)$, the centralizer of $e$, namely, such that $(q^j| q_i)=\delta_{ij}$.
As shown in \cite{structure},
the generators $w_i$, $i=1,\dots,\ell$, of the classical $\mc W$-algebra $\mc W(\mf g,f)$
can be uniquely chosen to have representatives $\widetilde{w}_i\in\widetilde{\mc W}$
such that $\widetilde{w}_i-q_i\in\langle[e,\mf g]\rangle$, the differential ideal of $\mc V(\mf g)$
generated by the subspace $[e,\mf g]\subset\mf g$.
We let
$$
q=\sum_{i=1}^\ell w_i\otimes q^i
\,\in\mc W\otimes\mf z(e)\,.
$$
Using \eqref{toprove} one can show that
there exist unique formal Laurent series $U(z)\in\mc W\otimes\mf{h}^\perp_{>0}$
and $h(z)\in\mc W\otimes\mf{h}_{>-1}$ such that
$$
e^{\ad U(z)}(\partial+1\otimes(f+zE)+q)=\partial+1\otimes (f+zE)+h(z)\,.
$$
The Laurent series $h(z)$ is the core of integrability. Indeed,
let $\mf c(\mf h)$ be the center of $\mf h$, and let
$$
\mf a=\big\{a\in\mf c(\mf h)\,\big|\, [a,f_2]=0\big\}\subset\mf g((z^{-1}))
\,.
$$
For $a\in\mf a$, let
\begin{equation}\label{20200619:eq2}
\tint g_a(z)=\tint(1\otimes a|h(z))=\sum_{n\in\mb Z_{\geq 0}}\tint g_{a,n}z^{N-n}\,,
\end{equation}
where $N$ is the largest power of $z$ appearing in $\tint(1\otimes a|h(z))$
with non-zero coefficient,
and $(\cdot\,|\,\cdot)$ is defined in \eqref{20200625:eq1}.
Then 
$$
\mc A=\Span\{\tint g_{a,n}\mid a\in\mf a, n\in\mb Z_{\geq 0}\}\subset\mc W/\partial\mc W
$$
is an abelian subalgebra of
$\mc W/\partial\mc W$
with respect to both $0$ and $\infty$-Lie brackets,
defining a hierarchy of bi-Hamiltonian equations
$$
\frac{dw}{dt_{a,n}}
=\{{\tint g_{a,n}}_{\lambda}w\}_{0}
=\{{\tint g_{a,n+1}}_{\lambda}w\}_{\infty}
\,,
\quad w\in\mc W\,,a\in\mf a\,,n\in\mb Z_{\geq 0}\,.
$$
Moreover, 
if $a\in\mf a$ is not central in $\mf g((z^{-1}))$,
then $\dim\Span\{\tint g_{a,n}\}_{n\in\mb Z_{\geq0}}=\infty$.
Consequently, the above hierarchy is integrable, see \cite{DSKV13a,mamuka1}.

When $f=0$, the corresponding $\mc W$-algebra is just $\mc V(\mf g)$ (since the $\mf{sl}_2$-triple $\mf s$ is trivial). Choosing $E\in\mf g$ to be any semisimple element one gets the so-called \emph{homogeneous DS hierarchy}, see \cite{DSKV13a}. For $\mf g=\mf{gl}_N$, let $E$ be a diagonal matrix with distinct eigenvalues $s_i\in\mb F$, and let $a$ be a diagonal matrix with eigenvalues $a_i\in\mb F$, then
the first two equations of this hierarchy are ($1\leq i,j\leq N$):
$$
\frac{d u_{ij}}{dt_0}=(a_i-a_j)u_{ij},
\qquad
\frac{d u_{ij}}{dt_1}
=\gamma_{ij}u_{ij}^\prime
+\sum_{k}(\gamma_{ik}-\gamma_{kj})u_{ik}u_{kj}
\,,
$$
where
$u_{ij}$ are differential variables,
$\gamma_{ij}=\frac{a_i-a_j}{s_i-s_j}$ for $i\neq j$ and $\gamma_{ij}=0$ for $i=j$.
The last equation is known as the \emph{$N$-wave equation}.
For $\mf g=\mf{sl}_2$, these two equations are trivial, but the third one is non-trivial, and the Dirac reduction
of $\mc V(\mf g)$ described at the end of Section \ref{sec:8}, produces the NLS equation.

Let $\mf g=\mf{sl}_3$. We have two non-zero conjugacy classes: the principal and 
the minimal.
In the principal case, the corresponding classical affine $\mc W$-algebra (with $E$ a highest root vector) is the algebra of differential polynomials
in two variables $u_1$ and $u_2$ with $\lambda$-brackets
\begin{align*}
\{{u_1}_\lambda u_1\}_{z}&=(2\lambda+\partial)u_1-2\lambda^3\,,
\{{u_1}_\lambda u_2\}_{z}
=(3\lambda+\partial)u_2+3z\lambda\,,
\\
\{{u_2}_\lambda u_2\}_{z}
&=
\frac13(2\lambda+\partial)u_1^2
-\frac16(\lambda+\partial)^3u_1-\frac16\lambda^3u_1
-\frac14\lambda(\lambda+\partial)(2\lambda+\partial)u_1+\frac16\lambda^5\,.
\end{align*}
Letting $a=(f+zE)^2$, from equation \eqref{20200619:eq2}
one  gets $\tint g_0=\tint u_2$
and the corresponding Hamiltonian equation is
$$
\left\{\begin{array}{l}
\frac{d u_1}{dt}=2{u_2}^\prime\\
\frac{d u_2}{dt}=-\frac16u_1^{\prime\prime\prime}+\frac23u_1u_1^\prime\,.
\end{array}\right.
$$
Eliminating ${u_2}$ from the system we get that $u=u_1$ satisfies the
\emph{Boussinesq equation}
$$
u_{tt}=-\frac{1}{3}u^{(4)}+\frac43(uu^\prime)^\prime\,.
$$
In the minimal case the classical affine $\mc W$-algebra (for the choice $\mf l=0$ and $E$ a highest root vector) is the algebra of differential polynomials in four
variables $u_1$, $u_2$, $u_3$ and $u_4$ with $\lambda$-brackets
\begin{align*}
\{{u_1}_\lambda {u_1}\}_{z}&=(2\lambda+\partial)u_1-\frac12\lambda^3+2z\lambda,\\
\{{u_1}_\lambda u_2\}_{z}&=(\frac32\lambda+\partial)u_2,\\
\{{u_1}_\lambda u_3\}_{z}&=(\frac32\lambda+\partial)u_3,\\
\{{u_1}_\lambda u_4\}_{z}&=(\lambda+\partial)u_4,\\
\{u_2{}_\lambda u_2\}_{z}&=0,\\
\{u_2{}_\lambda u_3\}_{z}&=
-u_1+\frac13u_4^2-\frac12(2\lambda+\partial)u_4+\lambda^2-z,\\
\{u_2{}_\lambda u_4\}_{z}&=3 u_2,\\
\{u_3{}_\lambda u_3\}_{z}&=0,\\
\{u_3{}_\lambda u_4\}_{z}&=-3 u_3,\\
\{u_4{}_\lambda u_4\}_{z}&=6 \lambda\,.
\end{align*}
%
Letting $a=f+zE$, from equation \eqref{20200619:eq2}
we get $\tint g_0=\tint u_4$ and $\tint g_1=\tint 6u_2 u_3$\,.
The corresponding first two Hamiltonian equations are
$$
\left\{\begin{array}{l}
\frac{d u_1}{dt_0}=0\\
\frac{d u_2}{dt_0}=-3u_2\\
\frac{d u_3}{dt_0}=3u_3\\
\frac{d u_4}{dt_0}=0
\end{array}\right.
\,,\quad
\left\{\begin{array}{l}
\frac{d u_1}{dt_1}=(u_2 u_3)'\\
\frac{d u_2}{dt_1}=-3u_2''+3u_1u_2-u_2 u_4^2-\frac32u_2 u_4'-3 u_2' u_4\\
\frac{d u_3}{dt_1}=3u_3''-3u_1u_3+u_3 u_4^2-\frac32u_3 u_4'-3 u_3' u_4\\
\frac{d u_4}{dt_1}=0\,.
\end{array}\right.
$$
Note that the differential variable $u_4$ does not evolve in time. Hence, we can perform a Dirac reduction
and get a Dirac reduced bi-Hamiltonian hierarchy whose first two equations are
$$
\left\{\begin{array}{l}
\frac{d u_1}{dt_0}=0\\
\frac{d u_2}{dt_0}=-3u_2\\
\frac{d u_3}{dt_0}=3u_3
\end{array}\right.
\,,\quad
\left\{\begin{array}{l}
\frac{d u_1}{dt_1}=(u_2 u_3)'\\
\frac{d u_2}{dt_1}=-3u_2''+3u_1u_2\\
\frac{d u_3}{dt_1}=3u_3''-3u_1u_3\,.
\end{array}\right.
$$
The second one is the well known Yajima-Oikawa (YO) equation \cite{YO76}.

\section{Double PVA and non-commutative Hamiltonian equations}
\label{sec:9}

In recent years there have been attempts to develop a theory of integrable  systems on non-commutative
associative algebras (see e.g. \cite{OS98,MS00,ORS13}).
An important advance in this direction was made by Van den Bergh, who in \cite{VdB08} introduced the
notion of a \emph{double Poisson algebra} structure
in a non-commutative associative algebra $V$.
His basic idea was to consider a $2$-fold bracket $\ldb-,-\rdb$ on $V$,
 with values in $V\otimes V$.
The Leibniz rules of a $2$-fold bracket are almost identical to the usual Leibniz rules:
\begin{align}
&
\ldb a,bc\rdb=\ldb a,b\rdb c+b\ldb a,c\rdb\,,
\label{eq:0.7}
\\
&\ldb ab,c\rdb=\ldb a,c\rdb\star b+a\star\ldb b,c\rdb\,,
\label{eq:0.8}
\end{align}
where 
\begin{equation}\label{eq:double-notation}
\begin{split}
& (a_1\otimes a_2) a_3=a_1\otimes (a_2a_3)
\,\,,\quad
a_1(a_2\otimes a_3)=(a_1a_2)\otimes a_3
\,,\\
& (a_1\otimes a_2)\star a_3=(a_1a_3)\otimes a_2
\,\,,\quad
a_1\star(a_2\otimes a_3)=a_2\otimes (a_1a_3)
\,.
\end{split}
\end{equation}
The skewsymmetry axiom is
\begin{equation}\label{eq:0.9}
\ldb a,b\rdb=-\ldb b,a\rdb^\sigma\,,
\end{equation}
where $\sigma$ is the permutation of factors in $V\otimes V$,
and the Jacobi identity is
\begin{equation}\label{eq:0.10}
\ldb a,\ldb b,c\rdb\rdb_L-\ldb b,\ldb a,c\rdb\rdb_R=\ldb\ldb a,b\rdb,c\rdb_L\,,
\end{equation}
where we denote $\ldb a_1, a_2\otimes a_3\rdb_L=\ldb a_1,a_2\rdb\otimes a_3$,
$\ldb a_1,a_2\otimes a_3\rdb_R=a_2\otimes\ldb a_1,a_3\rdb$,
$\ldb a_1\otimes a_2,a_3\rdb_L=\ldb a_1,a_3\rdb\otimes_1a_2$,
and $(a\otimes b)\otimes_1c=a\otimes c\otimes b$.

An associative algebra $V$, endowed with a $2$-fold bracket $\ldb-,-\rdb$ satisfying
axioms \eqref{eq:0.7}-\eqref{eq:0.10}, is called a \emph{double Poisson algebra}.

Given a double Poisson algebra $V$, Van den Bergh defines the following bracket on $V$ with values
in $V$:
\begin{equation}\label{eq:0.11}
\{a,b\}=\mult\ldb a,b\rdb\,,
\end{equation}
where $\mult:V\otimes V\to V$ is the multiplication map.
The bracket \eqref{eq:0.11} still satisfies the left Leibniz rule, but does not satisfy, in general,
other axioms of a Poisson bracket. However, this bracket induces well-defined
linear maps
$$
\quot{V}{[V,V]}\otimes V\to V
\qquad
\text{and}
\qquad
\quot{V}{[V,V]}\otimes \quot{V}{[V,V]}\to \quot{V}{[V,V]}\,,
$$
given by
\begin{equation}\label{eq:0.12}
\{\Tr(a),b\}=\{a,b\}
\qquad\text{and}\qquad
\{\Tr(a),\Tr(b)\}=\Tr\{a,b\}\,,
\end{equation}
where $\Tr:V\to\quot{V}{[V,V]}$ is the quotient map and $[V,V]$ is the linear span of commutators
$ab-ba$.
These maps have the following properties, important for the theory of non-commutative Hamiltonian
ODEs: the vector space $\quot{V}{[V,V]}$ is a Lie algebra and one has its representation on $V$ by derivations,
both defined by \eqref{eq:0.12}.

Brackets \eqref{eq:0.12} allow one to define the basic notions of a Hamiltonian theory.
Given a \emph{Hamiltonian function} $\Tr(h)$, $h\in V$, one defines the associated
\emph{Hamiltonian equation}
\begin{equation}\label{eq:0.13}
\frac{dx}{dt}=\{\Tr(h),x\}\,,
\quad x\in V\,.
\end{equation}
Two Hamiltonian functions $\Tr(f)$ and $\Tr(g)$ are said to be in involution if
\begin{equation}\label{eq:0.14}
\{\Tr(f),\Tr(g)\}=0\,.
\end{equation}

The notion of a double PVA introduced in \cite{DSKV15a} is 
obtained by merging the notions of
a double Poisson algebra and of a PVA. Let $\mc V$ be a
unital associative algebra with a derivation $\partial$.
We define a \emph{$2$-fold $\lambda$-bracket} on $\mc V$ as a map
$\ldb-_\lambda-\rdb:\mc V\otimes\mc V\to(\mc V\otimes\mc V)[\lambda]$,
satisfying sesquilinearity and left and right Leibniz rules
similar to \eqref{sesqui} and \eqref{lleibniz},\eqref{rleibniz} for PVA.
More precisely, the sequilinearity axioms become
\begin{equation}\label{20140702:eq4b}
\ldb \partial a_\lambda b\rdb=-\lambda \ldb a_\lambda b\rdb
\,\,,\,\,\,\,
\ldb a_\lambda \partial b\rdb=(\lambda+\partial) \ldb a_\lambda b\rdb
\,,
\end{equation}
while the left and right Leibniz rules can be written, respectively, as:
\begin{equation}\label{20140702:eq6b}
\begin{array}{l}
\displaystyle{
\vphantom{\Big(}
\ldb a_\lambda bc\rdb
=
\ldb a_\lambda b\rdb c+b\ldb a_\lambda c\rdb
\,} \\
\displaystyle{
\vphantom{\Big(}
\ldb ab_\lambda c\rdb
=
\ldb a_{\lambda+x} c\rdb \star(|_{x=\partial} b)
+(|_{x=\partial}a)\star\ldb b_{\lambda+x} c\rdb
\,,}
\end{array}
\end{equation}
where we are using the notation \eqref{eq:double-notation}
and, for $p(\lambda)=\sum_np_n\lambda^n\in (\mc V\otimes \mc V)[\lambda]$ and $v\in\mc V$,
we denote $p(\lambda+x)\star(|_{x=\partial}v)=\sum_np_n\star((\lambda+\partial)^nv)$,
and similarly for $(|_{x=\partial}v)\star p(\lambda+x)=\sum_n((\lambda+\partial)^nv)\star p_n$.

A \emph{double PVA} is the algebra $\mc V$, endowed with a $2$-fold $\lambda$-bracket
satisfying skewsymmetry  and Jacobi identity, which can be obtained in analogy 
with \eqref{skewsim} and \eqref{eq:0.9} (skewsymmetry),
and with \eqref{jacobi} and \eqref{eq:0.10} (Jacobi identity).
More precisely, the skewsymmetry axiom reads
\begin{equation}\label{eq:skew2}
\ldb a_{\lambda}b\rdb=-\ldb b_{-\lambda-\partial}a\rdb^\sigma\,,
\end{equation}
where $-\lambda-\partial$ in the RHS is moved to the left, acting on the coefficients.
The Jacobi identity can be written as 
\begin{equation}\label{eq:jacobi2}
\ldb a_{\lambda}\ldb b{}_\mu c\rdb\rdb_L
-\ldb b_{\mu}\ldb a{}_\lambda c\rdb\rdb_R
=\ldb\ldb a_{\lambda} b\rdb_{\lambda+\mu} c\rdb_L
\,,
\end{equation}
where we use the following notation.
For $a,b,c\in\mc V$,
\begin{align*}
& \ldb a_\lambda (b\otimes c)\rdb_L:=\ldb a_{\lambda} b\rdb\otimes c
\,\,,\,\,\,\,
\ldb a_\lambda (b\otimes c)\rdb_R:=b\otimes \ldb a_{\lambda} c\rdb
\,, \\
& \ldb (a\otimes b)_\lambda c\rdb_L
:=
\ldb a_{\lambda+x} c\rdb\otimes_1 (|_{x=\partial}b)
\,.
\end{align*}

If $\partial =0$, the axioms of a double PVA turn into the axioms of a double Poisson algebra if we let $\lambda=\mu=0$.


For a double PVA $\mc V$, we denote by $\tint$ the quotient map
$\mc V\to\overline{\mc V}:=\quot{\mc V}{([\mc V,\mc V]+\partial\mc V)}$.
Similarly to \eqref{eq:0.11}, define the map $\{-_\lambda-\}:\mc V\otimes\mc V\to \mc V[\lambda]$ by
\begin{equation}\label{eq:0.23}
\{a_\lambda b\}=\mult\ldb a_\lambda b\rdb
\,.
\end{equation}
As for the PVA, \eqref{eq:0.23} induces well-defined maps
$$
\overline{\mc V}\otimes\mc V\to\mc V
\quad\text{and}\quad
\overline{\mc V}\otimes\overline{\mc V}\to\overline{\mc V}
\,,
$$
given by formulas, similar to \eqref{eq:0.12}:
\begin{equation}\label{eq:0.24}
\{\tint a, b\}=\{a_\lambda b\}\big|_{\lambda=0}
\quad\text{and}\quad
\{\tint a,\tint b\}=\tint\{a_\lambda b\}\big|_{\lambda=0}
\,.
\end{equation}
As before, these maps have properties important for the theory of non-commutative
Hamiltonian PDEs: $\overline{\mc V}$ is a Lie algebra and one has its representation on $\mc V$
by derivations, both defined by \eqref{eq:0.24}.
We have definitions of a Hamiltonian function $\tint h\in\overline{\mc V}$, the corresponding
Hamiltonian equation in $\mc V$, similar to \eqref{eq:0.13},
and involutiveness, similar to \eqref{eq:0.14}, where $\Tr$ is replaced by $\tint$.

We define an \emph{algebra of (non-commutative) differential functions} (cf. \cite[Sec. 1]{BDSK09})
as a unital associative algebra $\mc V$ with a derivation $\partial$ and ``strongly commuting'' $2$-fold derivations
$\frac{\partial}{\partial u_i^{(n)}}:\,\mc V\to\mc V\otimes\mc V$, 
$i=1,\dots,\ell$, $n\in\mb Z_{\geq0}$, 
such that the following two properties hold:
%
for each $f\in\mc V$,
$\frac{\partial f}{\partial u_i^{(n)}}=0$
for all but finitely many $(i,n)$, and 
$$
\left[\frac{\partial}{\partial u_i^{(n)}},\partial\right]=\frac{\partial}{\partial u_i^{(n-1)}}\,.
$$
(Here, $\partial$ is extended to a derivation of $\mc V\otimes\mc V$ by letting
$\partial(a_1\otimes a_2)=(\partial a_1)\otimes a_2+a_1\otimes(\partial a_2)$.)
We say that the $2$-fold derivations $D$ and $E$
\emph{strongly commute} if $D_L\circ E=E_R\circ D$, where
$D_L(a_1\otimes a_2)=(Da_1)\otimes a_2$ and
$D_R(a_1\otimes a_2)=a_1\otimes(Da_2)$.

The most important example of an algebra of non-commutative
differential functions is the algebra of non-commutative differential polynomials
$\mc R_\ell$ in the indeterminates $u_i^{(n)}$, $i=1,\dots,\ell$, $n\in\mb Z_{\geq0}$, with the
derivation $\partial$ defined by $\partial u_i^{(n)}=u_i^{(n+1)}$,
with partial derivatives $\frac{\partial}{\partial u_i^{(n)}}$
defined on generators as 
$\frac{\partial}{\partial u_i^{(n)}}u_j^{(m)}=\delta_{i,j}\delta_{m,n}\,1\otimes1$
and extended to $\mc R_\ell$ by the Leibniz rule.
Any $2$-fold $\lambda$-bracket on $\mc R_\ell$ is given by the following Master Formula:
\begin{equation}\label{eq:0.27}
\ldb f_{\lambda}g\rdb
=\sum_{\substack{i,j\in I\\m,n\in\mb Z_{\geq0}}}
\frac{\partial g}{\partial u_j^{(n)}}
\bullet
(\lambda+\partial)^n
H_{ij}(\lambda+\partial)
(-\lambda-\partial)^m
\bullet
\left(\frac{\partial f}{\partial u_i^{(m)}}\right)^\sigma\,,
\end{equation}
where $H_{ij}(\lambda)=\ldb u_j{}_\lambda u_i\rdb\in(\mc R_\ell\otimes\mc R_\ell)[\lambda]$
and $\bullet$ is the associative product on $\mc V\otimes\mc V$
defined by $(a\otimes b)\bullet(c\otimes d)=(ac)\otimes(db)$. 
Formula \eqref{eq:0.27} defines a $2$-fold $\lambda$-bracket on any non-commutative algebra of differential functions
$\mc V$. This $2$-fold $\lambda$-bracket is skewsymmetric if and only if
$H(\partial)=\left(H_{ij}(\partial)\right)_{i,j=1}^\ell$ is a skewadjoint differential operator over
$\mc V\otimes\mc V$ and, provided that $H(\partial)$ is skewadjoint, the Jacobi identity holds
if and only if it holds on any triple of generators, see \cite[Sec. 3.3]{DSKV15a}.

The non-commutative Hamiltonian PDE, associated to the matrix differential operator
$H(\partial)$, defining the $2$-fold $\lambda$-bracket, and to the Hamiltionan functional
$\tint h\in\overline{\mc V}$, is the following evolution PDE, where $u=(u_i)_{i=1}^\ell$:
$$
\frac{du}{dt}
=\mult\Big(H(\partial)\bullet\Big(\frac{\delta h}{\delta u}\Big)^\sigma\Big)
\,,
$$
where $\frac{\delta h}{\delta u}
=\Big(\frac{\delta h}{\delta u_{i}}\Big)_{i=1}^\ell\in(\mc V\otimes\mc V)^{\ell}$
is the vector of $2$-fold variational derivatives
$$
\frac{\delta h}{\delta u_{i}}
=\sum_{n\in\mb Z_{\geq0}}(-\partial)^n\frac{\partial h}{\partial u_{i}^{(n)}}\,.
$$
The corresponding Lie bracket $\{-,-\}$ on $\overline{\mc V}$ is given by:
$$
\{\tint f,\tint g\}
=\tint \mult\left(\frac{\delta g}{\delta u}\bullet H(\partial)\bullet \Big(\frac{\delta f}{\delta u}\Big)^\sigma
\right)\,.
$$
The main motivation for introducing double Poisson algebras in \cite{VdB08} was the observation that, 
given a double Poisson algebra structure $\ldb\cdot\,,\,\cdot\rdb$
on a unital finitely generated associative algebra $V$, and 
a positive integer $N$,  we have an associated 
Poisson algebra structure on the algebra $V_N$ of polynomial functions on
the affine algebraic variety of $N$-dimensional representations of the algebra $V$ .
Explicitly, the algebra $V_N$ is isomorphic to the commutative
associative algebra with generators
$\{a_{ij}\mid a\in V, i,j=1,\dots,N\}$, subject to the relations ($k\in\mb F$, $a,b\in V$):
$(ka)_{ij}=ka_{ij},\,
(a+b)_{ij}=a_{ij}+b_{ij},\,
(ab)_{ij}=\sum_{k=1}^Na_{ik}b_{kj}$.
The Poisson bracket on the generators of $V_N$ is
$$
\{a_{ij},b_{hk}\}
=
\ldb a,b\rdb^\prime_{hj}\ldb a,b\rdb^{\prime\prime}_{ik}
\,,
$$
where we use Sweedler's notation $x=x^\prime\otimes x^{\prime\prime}$
for elements in $V\otimes V$.

In a similar manner, if $\mc V$ is a (non-commutative) differential associative algebra
with derivation $\partial$, then $\mc V_N$ is a
commutative differential associative algebra with derivation defined
on generators by $\partial(a_{ij})=(\partial a)_{ij}$.
It is then proved in \cite[Sec. 3.7]{DSKV15a} that, given a double PVA structure on $\mc V$, one can associate
for each positive integer $N$ a PVA structure on $\mc V_N$.

The simplest example of a double PVA is the following family of compatible double PVA structures
on $\mc R_1=\mb F\langle u, u', u'',\dots\rangle$:
$$
\ldb u_\lambda u\rdb=1\otimes u-u\otimes1+c(1\otimes1)\lambda\,,
$$
where $c\in\mb F$ is a parameter \cite[Sec.3.8]{DSKV15a}. In this case the associated PVA structure on
$(\mc R_1)_N=\mb F[u_{ij}^{(n)}\mid i,j=1,\dots,N,n\in\mb Z_{\geq0}]$
is just the affine PVA for $\mf{gl}_N$.
As an application,  double PVA are used to 
prove integrability of the non-commutative KP equation
$$
\left\{
\begin{array}{l}
\displaystyle{
u_y=w'\,,
}\\
\displaystyle{
3w_y=4u_t-u'''-3(u^2)'+3[u,w]
\,,}
\end{array}\right.
$$
and of the non-commutative $n$-th Gelfand-Dickey equation.
For $n=2$ this is the non-commutative KdV equation
$$
\frac{du}{dt}=\frac14\left(
u'''+3uu'+3u'u
\right)\,.
$$

\section{Multiplicative PVA and differential-difference equations}
\label{sec:10}

According to Kupershmidt's philosophy \cite{Kup85},
many ideas of the theory of integrable PDE should be extended to the theory
of integrable differential-difference equations.
It was observed in \cite{DSKVW18} that, in order to extend the ideas of the PVA theory to the theory 
of integrable Hamiltonian differential-difference equations, one is led to a ``multiplicative'' version of PVA.

By definition, a \emph{multiplicative PVA} (mPVA) is a unital commutative associative
algebra $\mc V$ with an automorphism $S$,
endowed with a multiplicative $\lambda$-bracket
$$
\mc V\otimes\mc V\,\rightarrow\,\mc V[\lambda,\lambda^{-1}]
\,\,,\,\,\,\,
a\otimes b\mapsto\{a_\lambda b\}
\,,
$$
such that the same left Leibniz rule \eqref{lleibniz} holds as in the ``additive'' case,
but we have the following multiplicative analogues of \eqref{sesqui},\eqref{skewsim},\eqref{jacobi}:
\begin{enumerate}
\item[M1]
(sesquilinearity)
$\{S(a)_\lambda b\}
=
\lambda^{-1}\{a_\lambda b\}$,
$\{a_\lambda S(b)\}
=
\lambda S\{a_\lambda b\}$;
\item[M2]
(skewsymmetry)
$\{b_\lambda a\}
=
-\{a_{\lambda^{-1}S^{-1}}b\}$;
\item[M3]
(Jacobi identity)
$\{a_\lambda\{b_\mu c\}\}-\{b_\mu\{a_\lambda c\}\}
=
\{\{a_\lambda b\}_{\lambda\mu}c\}$.
\end{enumerate}
Note that the left Leibniz rule and axiom M2 imply the right Leibniz rule
$$\{ab_\lambda c\}
=
\{a_{\lambda S}c\}_\to b+\{b_{\lambda S}c\}_\to a\,.
$$
(As usual, the arrow indicates that $S$ should be moved to the right.)

Note that, a mPVA carries the structure of a lattice Poisson algebra (= a Poisson algebra with an automorphism, which we denote by $S$) by
$$
\{a,b\}=a_{0}b\,,
$$
the coefficient of $\lambda^0$ in the $\lambda$-bracket $\{a_\lambda b\}$. Conversely, if $\mc V$ is a lattice Poisson algebra such that $\{S^n(a),b\}=0$ for all but finitely many $n\in\mb Z$,
then $\mc V$ is a mPVA with $\lambda$-bracket given by ($a,b\in\mc V$)
$$
\{a_\lambda b\}=\sum_{n\in\mb Z}\{S^n(a),b\}\lambda^n\,.
$$
A mPVA $ \mathcal{V} $ gives rise to a Hamiltonian differential-difference equation as follows. Denote by 
\[ \tint : \mc V \rightarrow \bar{\mc V} := \mc V / (S-1) \mc V \]
\noindent the canonical quotient map. Then, formula
\begin{equation}\label{e1.11}
\{ \tint f, \tint g   \} = \tint \{f_\lambda g \} |_{\lambda = 1}
\end{equation}
\noindent endows $ \bar{\mc V} $ with a Lie algebra structure, and the formula
$$
\{ \tint f, g   \} =  \{f_\lambda g \} |_{\lambda = 1}
$$
defines a representation of the Lie algebra $ \bar{\mc V} $ by derivations of the multiplicative PVA $ \mc V, $ which commute with $ S $.

Choosing a Hamiltonian functional $ \int h \in \bar{\mc V}, $ we define the corresponding Hamiltonian equation
\begin{equation}\label{e1.13}
\frac{du}{dt} = \{ \tint h, u \}, \ u \in \mc V. 
\end{equation}
A Hamiltonian function $ \tint h_1 $ is called an integral of motion of this equation if $ \tint \frac{dh_1}{dt} = 0 $ in virtue of \eqref{e1.13}, i.e. $ \{ \tint h, \tint h_1 \} = 0. $ The equation \eqref{e1.13} is called \emph{integrable} if it has infinitely many integrals of motion in involution, i.e. if $ \int h $ is contained in an infinite-dimensional abelian subalgebra of the Lie algebra $ \bar{\mc V} $ with bracket \eqref{e1.11}.

The most famous example of a differential-difference equation is the Volterra chain:
\begin{equation}\label{e1.14}
\frac{du}{dt} 
=
u(S(u) - S^{-1}(u))
\end{equation}
(applying $ S^n $ to both sides, we obtain its more traditional form
$$\frac{du_n}{dt} = u_n (u_{n+1} - u_{n-1}), \ n \in \mb Z\,,$$
where $u_n=S^n(u)$).
This equation can  be written in a Hamiltonian form \eqref{e1.13}, with respect to
the multiplicative $\lambda$-bracket defined on generators of the algebra of polynomials $\mc V=\mb F[S^n(u)\mid n\in\mb Z]$
by
$$
 \{ u_\lambda u \} =  u\lambda S(u) -u\lambda^{-1} S^{-1}(u) \,,
 $$
and to the Hamiltonian functional $\tint u $.
Moreover, equation \eqref{e1.14} is bi-Hamiltonian, i.e. it can be written in the form \eqref{e1.13} with a different
mPVA $\lambda$-bracket, called complementary type order 2 multiplicative Poisson $ \lambda $-bracket $ \{ u_\lambda u \}_{2, u, -1} $ 
in \cite{DSKVW18} and the Hamiltonian functional $\frac12 \tint \log u $.
These two multiplicative Poisson $ \lambda $-brackets are compatible. By applying the Lenard-Magri type scheme it is shown in \cite[Sec. 7]{DSKVW18} that the Volterra chain \eqref{e1.14} has infinitely many linearly independent integrals of motion, i.e. it is integrable.

The non-local mPVA are indispensable for the theory of integrable
Hamiltonian differential-difference equations as well.
While in the ``additive'' PVA case
the $\lambda$-brackets could be allowed to take values only in the Laurent series,
the ``multiplicative'' $\lambda$-brackets can be any bilateral series in $\lambda$.
However, for the ``multiplicative'' Dirac reduction one needs the $\lambda$-brackets to be rational,
i.e. symbols of rational difference operators. Details can be found in \cite[Sec. 5]{DSKVW19}.

The simplest example of a non-local mPVA on $\mc V=\mb F[S^n(u)\mid n\in\mb Z]$ is given by
\begin{equation}\label{eq:FR1}
\{u_\lambda u\}
=
u\frac{\lambda S-1}{\lambda S+1}u
\,.
\end{equation}
This $\lambda$-bracket is compatible with 
\begin{equation}\label{eq:FR2}
\{u_\lambda u\}=\lambda-\lambda^{-1}
\,.
\end{equation}
The $q$-deformed $\mc W_N$-algebras of Frenkel and Reshetikhin \cite{FR96} can be interpreted 
in the language of multiplicative PVA,
in particular, $\mc W_2$ corresponds to the sum of the two $\lambda$-brackets
\eqref{eq:FR1} and \eqref{eq:FR2}.
Applying the Lenard-Magri scheme to this pair,
one can prove integrability of the modified Volterra lattice (see \cite[Sec. 4]{DSKVW19})
$$
\frac{dv}{dt}
=
v^2(S^{-1}(v)-S(v))
\,.
$$
A convenient framework to study non-commutative differential-difference equations is provided by double mPVA, see \cite{CW,FV22}.

\section{Conclusion}
\label{sec:13}
In this article we explained the foundations of the theory of PVA, and their applications to integrability of Hamiltonian PDE.
We also touched upon the theory of double PVA and multiplicative PVA along with their applications to integrability
of non-commutative PDE and of differential-difference equations respectively.

We hope that this article will call attention of experts in the theory of integrable systems to the method of $\lambda$-brackets,
central to the article.

\end{document}